\documentclass[12pt]{article}

\usepackage{epsfig}
\usepackage{a4}

\newcounter{subequation}[equation]

\makeatletter

\expandafter\let\expandafter\reset@font\csname reset@font\endcsname
\newenvironment{subeqnarray}
  {\arraycolsep1pt
    \def\@eqnnum\stepcounter##1{\stepcounter{subequation}{\reset@font\rm
      (\theequation\alph{subequation})}}\eqnarray}%
  {\endeqnarray\stepcounter{equation}}

\newcounter {subsubsubsection}[subsubsection]
\renewcommand\thesubsubsubsection{\thesubsubsection .\@arabic\c@subsubsubsection}
\newcommand\subsubsubsection{\@startsection{subsubsubsection}{4}{\z@}%
                                     {-3.25ex\@plus -1ex \@minus -.2ex}%
                                     {1.5ex \@plus .2ex}%
                                     {\normalfont\normalsize\bfseries}}
\let\subsubsubsectionmark\@gobble
\newcommand*\l@subsubsubsection{\@dottedtocline{4}{7.0em}{4.1em}}

\makeatother

\gdef\be#1\ee{\begin{equation}#1\end{equation}}
\gdef\bea#1\eea{{\arraycolsep1pt\begin{eqnarray}#1\end{eqnarray}}}

\gdef\bsea#1\esea{\begin{subeqnarray}#1\end{subeqnarray}}

\newcommand{\bu}{{\bar u}}
\newcommand{\bh}{{\bar h}}
\newcommand{\bv}{{\bar v}}

\newcommand{\tY}{{\tilde Y}}
\newcommand{\tr}{{\tilde r}}
\newcommand{\tn}{{\tilde n}}
\newcommand{\tk}{{\tilde\kappa}}
\newcommand{\tw}{{\tilde w}}
\newcommand{\tu}{{\tilde u}}
\newcommand{\tv}{{\tilde v}}
\newcommand{\ty}{{\tilde y}}
\newcommand{\tg}{{\tilde\gamma}}
\renewcommand{\th}{{\tilde h}}

\newcommand{\NoAlign}[1]{\noalign{\vskip\belowdisplayskip
  \noindent#1\par\vskip\abovedisplayskip}}

\begin{document}

\date{}

\title{\bf Gravitating BPS Monopoles in all $d=4p$ Spacetime Dimensions}
\author{{\large Peter Breitenlohner}$^{\ddagger}$
and {\large D. H. Tchrakian}$^{\star \dagger}$ \\ \\
$^{\ddagger}${\small Max-Planck-Institut f\"ur Physik}\\
{\small Werner-Heisenberg-Institut}\\
{\small F\"ohringer Ring 6, D-80805 M\"unchen, Germany}\\ \\
$^{\star}${\small School of Theoretical Physics,
Dublin Institute for Advanced Studies,} \\
{\small 10 Burlington Road, Dublin 4, Ireland} \\
$^{\dagger}${\small Department of Computer Science,
National University of Ireland Maynooth,} \\
{\small Maynooth, Ireland}}


\maketitle


\begin{abstract}

We have constructed, numerically, both regular and black hole static
solutions to the simplest possible gravitating Yang-Mills--Higgs (YMH) in
$4p$ spacetime dimensions.  The YMH systems consist of $2p-$th power
curvature fields without a Higgs potential.  The gravitational systems
consist of the `Ricci scalar' of the $p-$th power of the Riemann curvature. 
In $4$ spacetime dimensions this is the usual Einstein-YMH (EYMH) studied in
\cite{Breitenlohner:1991aa, Breitenlohner:1994di}, whose qualitative results
we emulate exactly.

\end{abstract}

\section{Introduction}
Gravitating monopoles were studied intensively in
\cite{Breitenlohner:1991aa, Breitenlohner:1994di, Lee:1991vy}, subsequent to
the discovery of the finite energy regular solutions~\cite{Bartnik:1988am}
to the gravitating Yang--Mills (EYM) system.  These regular EYM solutions were
soon extended to black holes~\cite{Volkov:1989fi}.  EYM solutions in $3+1$
dimensional spacetime are reviewed in \cite{Volkov:1998cc} and an exhaustive
analysis is given in \cite{Breitenlohner:1993es}.  Along with the
introduction of the (dimensionful) Higgs field that leads to the
construction of regular and black monopoles \cite{Breitenlohner:1991aa,
Breitenlohner:1994di, Lee:1991vy} in the gravitating EYM-Higgs (EYMH)
system, the (dimensionful) cosmological constant was introduced.  The case
of positive cosmological constant is reviewed quite adequately in
\cite{Volkov:1998cc} and that of negative is in the review
\cite{Winstanley:2008ac}.  The latter led to very interesting new features
of the solutions and is, in some sense, an alternative to the Higgs field.

EYM solutions in (higher) dimensions $D+1$ ($D\ge 4$) were considered only
relatively recently.  Here there are two possibilities of identifying the
$D$ spacelike dimensions: Either as asymptotically Minkowski with
$S^{D-1}$ boundary, or, with boundary $S^{D-N-1}\times R^{N}$ in which case
the $N$ codimensions are frozen as in the case of the $z-$coordinate of the
Abrikosov-Nielsen-Olesen vortex.  In the case of $S^{3}$ boundary for $D=4$
it was found in \cite{Volkov:2001tb} that the energy of the regular solution
is infinite, and in \cite{Okuyama:2002mh} (with negative cosmological
constant) it turned out that the energy of the black hole is also infinite. 
In addition, in \cite{Volkov:2001tb} solutions with boundary $S^{2}\times
R^1$ $D=4$ with finite energy {\it per unit} length along $R^1$ were
constructed, the total energy remaining infinite.  This is not surprising
because the usual EYM system in $D\ge 4$ does not have the requisite scaling
properties for there to exist finite energy solutions.  This obstacle is
circumvented by extending the definition of the EYM system to feature higher
order YM curvature terms with the appropriate scaling.

Suitably defined EYM systems featuring higher order YM curvature%
\footnote{Higher order Riemann curvature terms can also be introduced, but
since all studies are in practice carried out for systems subject to
symmetries, such terms of high enough order Riemann curvature vanish due to
the symmetry imposed.  Their inclusion is thus unnecessary.  Any
nonvanishing such terms included do not play an essential role, but rather
result only in a quantitative difference.}
terms in all $d=D+1$ spacetime dimensions were studied in
\cite{Brihaye:2002hr, Brihaye:2002jg} without cosmological constant, and in
\cite{Radu:2005mj, Brihaye:2007tw} with cosmological constant.  These latter
differ from their $d=3+1$ dimensional EYM analogues~\cite{Bartnik:1988am,
Volkov:1998cc}, which are unstable (sphaleronic) and have no gravity
decoupling limits, in that some of them do have gravity decoupling limits
which are stabilised by Chern--Pontryagin (instantonic) charges.  In this
respect these are more akin to the gravitating monopoles, and like them
these systems exhibit one or more dimensionful constants that cannot be
scaled away and hence parametrise the ensuing solutions.  As a result they
feature certain bifurcation properties like the monopoles but some of these
exhibit in addition what were named {\it conical singularities} in
\cite{Breitenlohner:2005hx}, where this analysis was carried out.  To date
no gravitating monopoles in dimensions higher than $d=3+1$ are constructed
and it is the aim of the present work to do that, in the simplest class of
YMH models defined in $4p$ spacetime dimensions.  Before going into the
details of this choice, we point out that this is a very interesting
restriction since in our previous study of higher dimensional EYM solutions
in \cite{Breitenlohner:2005hx}, it was found that the qualitative features
of these repeated {\it modulo} every $4-$dimensions.  Like in all the work
on gravitating gauge fields quoted above, our solutions take into account
the backreaction of gravity on the YMH fields.

Before considering the possible choices of suitable higher dimensional YMH
systems, it is in order to consider the particular interest in gravitating
monopoles in higher dimensions, {\bf a)} on a technical level, and {\bf b)}
from the viewpoint of applications.

{\bf a)} On a technical level, it is the only way which enables the
construction of static solutions of EYM systems in spacetime dimensions
higher than {\em four}, which describe a nonvanishing `non Abelian electric'
connection $A_0$.  In this respect EYM solutions in the presence of a
negative cosmological constant differ in spacetime dimension $d=4$, from
those in $d\ge 5$.  In $d=4$ the presence of the negative cosmological
constant results in such asymptotic properties, which in contrast to the
asymptotically flat case, allow for nonvanishing electric field.  This is a
consequence of the fact that the `magnetic' connection of these solutions
can~\cite{Winstanley:2008ac}, among other possibilities, behave as a {\em
half pure gauge} just like a monopole.  We have verified that the
corresponding EYM systems (with negative cosmological constant) in $d\ge 5$
support `magnetic' connections that behave asymptotically as {\em pure
gauge}, like their asymptotically flat counterparts.  Hence they do not
support a nonvanishing electric field.

When a Higgs field is introduced however (irrespective of the presence of a
cosmological constant), the situation changes.  Already in flat space, EYMH
systems automatically support dyonic solutions, {\it e.g.}, in all even
dimensional spacetimes for models employed in \cite{Radu:2005rf}.  Clearly,
when the gravitational force is switched on, the `electric' field of the
resulting EYMH solution will persist.

{\bf b)} From the viewpoint of applications, like their higher dimensional
EYM analogues, they are expected to be relevant to various aspects of the
study of $D-$branes.  Some examples in the literature concern monopoles in
string theory~\cite{Gauntlett:1992nn, Bergshoeff:2006bs}, and
selfgravitating supersymmetric solitons in \cite{Gibbons:1993xt}.  In the
former examples~\cite{Gauntlett:1992nn, Bergshoeff:2006bs} only the
Yang-Mills system in the absence of Higgs fields appears, so that their
higher dimensional extensions will feature the gauge field configurations
described by the Yang-Mills hierarchy.%
\footnote{In \cite{Gauntlett:1992nn, Bergshoeff:2006bs} instanton fields of
the usual YM system are exploited, while their higher dimensional extensions
would employ the corresponding instantons of the YM hierarchy introduced in
\cite{Tchrakian:1984gq}.}
The work of \cite{Gibbons:1993xt} on the other hand concerns
monopoles of the usual YMH system in the {\it spacelike} subspace.  Thus its
higher dimensional extensions will involve field configurations of the
YM-Higgs hierarchies (see below), in other words higher dimensional monopoles. 
What is more, is that the monopoles in \cite{Gibbons:1993xt} satisfy the
selfduality constraint in the spacelike dimensions, and that the higher
dimensional extensions of the YMH systems employed here are chosen precisely
such that the corresponding extended Bogomol'nyi constraints are likewise in
force.

When choosing a YMH system in arbitrary (Euclidean) dimension $D$ supporting
a monopole solution we are faced with a plethora models.  The most efficient
way of constructing these YMH models is {\it via} dimensional
descent~\cite{O'Brien:1988xr} from a YM system~\cite{Tchrakian:1984gq} on
$D+N$ (even) dimension.  Integrating out the coordinates on the (compact)
$N$ codimensions results in the $D-$dimensional residual YMH theory
supporting monopoles, whose topological (monopole) charge is the descendent
of the $\frac12\,(D+N)-$th Chern--Pontryagin charge.  We shall not dwell on
the detailed properties of these $D-$dimensional monopoles here, save to
emphasise their most relevant feature pertinent to the present work. 
Because of the high degree of nonlinearity of the extended selfuality
equations~\cite{Tchrakian:1984gq} in $D+N$ dimensions, the descendent
Bogomol'nyi equations in $D$ dimensions are {\it
overdetermined}~\cite{Tchrakian:1990gc} and in general cannot be saturated. 
It turns out that these Bogomol'nyi equations can be saturated only when the
descent is over $N=1$ and $N=D-2$ codimensions, the latter case being
irrelevant here.  Since the Euclidean dimensions $D+N$ are only even, then
the residual Euclidean space is $D-1$ dimensional, restricting us to models
in even spacetime dimensions $d=(D-1)+1$ only.

In the present work we have restricted our attention to $d=4p$ dimensional
spacetimes because the YM hierarchy in $D+1=4p$ dimensions is scale
invariant and does not feature an additional dimensionful constant.%
\footnote{YM systems on $R^{2p+2}$ supporting instantons involve at least
one additional dimensionful constant (see {\it e.g.}, \cite{Burzlaff:1993kf})
whose dimensional descendent therefore will feature one more dimensional
constant in addition to the Higgs VEV.}
This results in the simplest possible residual YMH models, keeping a tight
analogy with the 't~Hooft--Polyakov monopole (in the BPS limit).  Like the
latter, these models feature only one dimensional constant, namely the Higgs
VEV with inverse dimension of a length.  A more direct construction, with
the sole criterion of achieving a topological lower bound and not applying
dimensional descent, was employed in \cite{Tchrakian:1978sf, Kihara:2004yz,
Radu:2005rf}, but these models always feature an extra dimensional constant
in addition to the Higgs VEV, considerably complicating the numerical
analysis in the gravitating case.  Thus we have eschewed such models here.

Finally, we state the action densities to be employed.  It is convenient to
express the connection and the Higgs field in terms of the chiral $so(4p)$
representation matrices
\be
\label{sig}
\Sigma_{\mu\nu}=(\Sigma_{ij},\Sigma_{i,D+1})\;,
\qquad
\mu=0,i\;,
\quad i=1,2,\dots,D\;,
\ee
such that the connection $A_i$ takes its values in the algebra of
$SO(4p-1)$, namely $\Sigma_{ij}$, and the Higgs field is
\be
\Phi=\phi^i\,\Sigma_{i,D+1}\;.
\ee
In terms of the curvature $F\equiv F(2)$ of $A$ and the covariant derivative
$D\Phi$, the action densities of this hierarchy of YMH models in $4p$
spacetime are defined as
\be\label{YMH}
{\cal S}_{\rm matter}^{(4p)}=\mbox{Tr}\,\left[\frac{1}{2\,(2p)!}F(2p)^2
-\frac{1}{2\,(2p-1)!}(F(2p-2)\wedge D\Phi)^2\right]\;,
\ee
the $2k-$form $F(2k)$ being the $k-$fold totally antisymmetrised product of
$F=F(2)$. The flat space static dyons of these systems were constructed in
\cite{Radu:2005rf}. It is the monopole solutions in \cite{Radu:2005rf} that
are gravitated in the present work, in the spirit of
\cite{Breitenlohner:1991aa, Breitenlohner:1994di}.

A pertinent comment at this point is that had we chosen the Higgs kinetic
term in (\ref{YMH}) to be $(F(2k-2)\wedge D\Phi)^2$, with $k\neq p$ in $2(p+k)$
dimensions, the solutions would still be topologically
stable~\cite{Radu:2005rf} but now a new dimensional constant will appear in
(\ref{YMH}). In this respect such systems would be more akin to the higher
dimensional models studied in \cite{Breitenlohner:2005hx} and their
gravitating solutions could then exhibit {\it conical fixed points}. The
fixed point analysis for such selfgravitating YMH has not been carried out
to date and it promises to be appreciably more involved than the
corresponding analysis in \cite{Breitenlohner:2005hx}, basically because now
there is a Higgs field function in addition to the gauge field function. We
have eschewed this choice here keeping strictly the analogy with the
gravitating monopoles in $3+1$ dimensions, and encounter no
solutions featuring {\it conical fixed points}. We will return to this
question elsewhere.

To complete the definition of the gravitating YMH systems, the gravitational
Lagrangian must be specified. Restricting to Levi--Civita connections, the
hierarchy of Einstein systems in $d=2(p+q)$ dimenional spacetimes are
\be\label{gp}
e\,R_{(p,q)}=\langle E(2p),R(2p)\rangle\;,
\ee
where $E(2p)={}^{\star}e(2q)^{\star}$ is the double--Hodge dual of $e(2q)$,
the $2q-$fold antisymmetrised product of the {\it Vielbein} fields.  $e$ in
(\ref{gp}) is the determinant of the {\it Vielbein} and $R(2p)$ is the
$2p-$form Riemann curvature.  The $p=1$ member of the hierarchy (\ref{gp})
is the Einstein--Hilbert Lagrangian in $d=2(q+1)$, $p=2$ the Gauss--Bonnet
system in $d=2(q+2)$, {\it etc.} Our choice of Lagrangian employed for
gravitating the YMH system (\ref{YMH}) for given $p$ is the
$p=q=\frac{d}{4}$ member of (\ref{gp}), like in the $p=1$ case studied in
\cite{Breitenlohner:1991aa, Breitenlohner:1994di}.  One can of course choose
the $p=1$ member of (\ref{gp}) with all YMH actions~(\ref{YMH}) but in that
case the relative dimensionality of the YMH and gravitational Lagrangians
would be $[L^{2(2p-1)}]$ instead of $[L^{2p}]$, which then would match with
the $p=q=1$ case studied in \cite{Breitenlohner:1991aa,
Breitenlohner:1994di}, only for $p=1$.  This would result in the dilution of
the analogy with the latter at least on the quantitative level.  This choice
was made previously in the study of the EYM systems in $4p$ dimensional
spacetimes in \cite{Radu:2006mb}, where it resulted in the tight similarity
in the qualitative properties of the solutions of the $p-$th YM system
gravitated with the $p(=q)-$th Einstein systems in all $4p$ dimensions.

The hierarchy of monopoles to be studied here is that of static and
spherically symmetric (in $4p-1$ space dimensions) solutions to the
equations arising from the action densities
\be\label{lag}
{\cal S}^{(4p)}=e\left(R_{(p,p)}+{\cal S}_{\rm matter}^{(4p)}\right)\,.
\ee

\section{Ansatz, Action, and Differential Equations}
Using the spherically symmetric metric Ansatz in `Schwarzschild'
coordinates
\be\label{Smetric}
ds^2=A^2(r)\mu(r)dt^2-\frac{dr^2}{\mu(r)}-r^2d\Omega^2_{(d-2)}\;,
\ee
together with a generalised 't~Hooft--Polyakov Ansatz for the magnetic
monopole
\be\label{connection}
A_i=\Sigma_{ij}\frac{x^j}{r^2}\Bigl(1-w(r)\Bigr)\;,
\qquad
\phi^i=\frac{x^i}{r}h(r)\;.
\ee
we obtain the reduced action
\be\label{reducedaction}
S=\frac{p(d-2)!}{(d-2p-1)!}(\tilde\kappa_pS_G+\tilde\tau_pS_M)\;,
\ee
with $\tilde\kappa_p=\kappa_p/2^{2(p-1)}$, $\tilde\tau_p=\tau_p/(2p)!$, and
\bsea\label{reducedactionGM}
S_G&=&-\frac{1}{2p}\int dr\,A\frac{d}{dr}\Bigl(r^{d-2p-1}(1-\mu)^p\Bigr)\;,
\\
S_M&=&\int dr\,A\,r^{d-4p}
  \Biggl[W^{p-1}\mu\Bigl(\frac{dw}{dr}\Bigr)^2
  +\frac{d-2p-1}{2p}\frac{W^p}{r^2}
\nonumber\\&&\qquad\qquad\qquad
  +\frac{\mu}{2p}\Bigl(r\frac{dH_p}{dr}\Bigr)^2
  +(d-2p-1)w^2H_p^2\Biggr]\;,
\esea
where $W=(w^2-1)^2$ and $H_p=(w^2-1)^{p-1}h$.

A Derrick-type scaling argument shows that static finite energy solutions of
the field eqs.\ derived form the action~(\ref{reducedaction}) can only exist in
spacetime dimensions $d$ with
\be
2p+1<d<4p+1\;,
\ee
and indeed in \cite{Radu:2006mb} the EYM system has been studied for all
values in this range. The presence of the Higgs field, however, requires
$d=4p$.

The action~(\ref{reducedaction}) depends on three dimensionful parameters,
$\tilde\kappa_p$, $\tilde\tau_p$, and the vacuum expectation value
$\eta=\lim_{r\to\infty}h(r)$ of the Higgs field, with the dimensionless
ratio
\be
\alpha=\eta
  \left(\frac{\tilde\tau_p}{\tilde\kappa_p}\right)^{\frac{1}{2p}}\;.
\ee
Since an overall factor in front of the action has no effect on the eqs.~of
motion, we can rescale $r$ such that either
\be
S={\rm const.}\cdot(S_G+\alpha^{2p}S_M)\;,
\qquad
\eta=1\;,
\ee
resulting in $\alpha$-dependent eqs.~of motion with $\alpha$-independent
boundary conditions, or
\be
S={\rm const.}\cdot(S_G+S_M)\;,
\qquad
\eta=\alpha\;,
\ee
resulting in $\alpha$-independent eqs.~of motion with an $\alpha$-dependent
boundary condition.

These two formulations are clearly equivalent as long as $\alpha\ne0$. In
the limit $\alpha\to0$ they are, however, quite different. The first
alternative describes a YMH system in a fixed gravitational background,
either flat (no gravity) or a Schwarzschild type black hole. The second
alternative describes the (self gravitating) EYM system of
\cite{Radu:2006mb}.

In the following we will mostly use the second formulation, but will
occasionally refer to the first formulation as `unscaled variables'.

The dependence of the action~(\ref{reducedaction}) on $A(r)$ suggests to
define the `mass function'
\be\label{mass}
M(r)=\frac{4p+1-d}{2p}r^{d-2p-1}(1-\mu)^p\;,
\ee
that has a finite limit $M=\lim_{r\to\infty}M(r)$ for asymptotically flat
solutions with finite energy.

For the Reissner-Nordstr\"om solution with $w\equiv0$ and $h\equiv\alpha$
the mass function satisfies
\bsea
\frac{dM(r)}{dr}&=&\frac{d-2p-1}{2p}(4p+1-d)r^{-(4p+2-d)}\;,
\\
M(r)&=&M-\frac{d-2p-1}{2p}r^{-(4p+1-d)}\;.
\esea
The normalization of $M(r)$ in Equ.~(\ref{mass}) is chosen such that $M=1$
for the extremal Reissner-Nordstr\"om solution. The limit $\alpha\to0$ (in
the original, unscaled variables) yields the generalised Schwarzschild
solutions with constant $M(r)=M$, and thus
\be
\mu(r)=1-\left(\frac{2pM}{4p+1-d}\right)^{1/p}r^{-(d-2p-1)/p}\;.
\ee

\subsection{New Variables -- General}
The metric Ansatz~(\ref{Smetric}) in terms of Schwarzschild coordinates used
so far can not describe situations where the radius $r$ of $D-1$-spheres
first increases and then decreases, as in the solutions studied
in~\cite{Volkov:1996qj, Breitenlohner:2000rp, Forgacs:2005nt} with spaces of
spherical topology.  Even for asymptotically flat spaces with monotonically
increasing $r$, the equations of motion resulting from the
action~(\ref{reducedaction}) are singular when $\mu(r)=0$, {\it i.e.}, at a
horizon as well as for certain `critical' solutions with a double zero of
$\mu(r)$.  Following \cite{Breitenlohner:1994di}, we avoid this coordinate
singularity by the most general static, spherically symmetric metric Ansatz
\be\label{taumetric}
ds^2=e^{2\nu(\tau)}dt^2-e^{2\lambda(\tau)}d\tau^2-r^2(\tau)d\Omega^2_{(d-2)}\;,
\ee
with $r$ now a function of $\tau$, and substituting
\be\label{tausubst}
A\,dr=e^{\nu+\lambda}\,d\tau\;,
\qquad
\mu=(e^{-\lambda}\dot r)^2\;,
\ee
with the notation $\dot x=dx/d\tau$.

In order to obtain a system of first order differential equations (dynamical
system), we introduce additional variables $n$, $\kappa$, $u$, and
$V_p$
\bsea\label{deffirstford}
e^{-\lambda}\dot r&=&n\;,
\\
re^{-\lambda}\dot\nu&=&\kappa-n\;,
\\
e^{-\lambda}\dot w&=&u\;,
\\
re^{-\lambda}\dot H_p&=&V_p\;,
\esea
following the procedure in, {\it e.g.}, \cite{Breitenlohner:1994di,
Breitenlohner:2003qj}.

First we observe that the metric Ansatz~(\ref{taumetric}) is explicitly
invariant under reparametrisation of the new `radial' variable $\tau$. 
Consequently varying the action cannot result in a differential equation for
$\lambda$, and thus all derivatives of $e^{-\lambda}\dot r$ in the action
can be absorbed into a surface term that can be discarded.  Then we
introduce the new variables $n$, $\kappa$, $u$, and $V_p$ as Lagrange
multipliers such that the resulting action contains at most one
$\tau$-derivative and variation w.r.t.\ to the new variables yields
Eqs.~(\ref{deffirstford}) as field equations.  Finally we discard yet
another surface term to obtain a compact expression.

Each of these three steps is straightforward for $p=1$, but requires some
new techniques for $p>1$ (see Appendix~A for the details). As a result
Eqs.~(\ref{reducedactionGM}) are expressed as
\bsea\label{Stau3}
S_G&=&\int d\tau\Bigl[e^{\nu+\lambda}\,r^{d-2p-2}(1-n^2)^{p-1}\Bigl(
(\kappa-n)n-\frac{d-2p-1}{2p}(1-n^2)
\nonumber\\&&\qquad
  +(re^{-\lambda}\dot n)
  -(\kappa-n)(e^{-\lambda}\dot r)\Bigr)\Bigr]\;,
\\
S_M&=&\int d\tau\,e^{\nu+\lambda}\,r^{d-4p}\Bigl[
(d-2p-1)\Bigl(\frac{1}{2p}\frac{W^p}{r^2}+w^2H_p^2\Bigr)
  -W^{p-1}u^2-\frac{V_p^2}{2p}
\nonumber\\&&\qquad\qquad\qquad
+2\Bigl(W^{p-1}u(e^{-\lambda}\dot w)
  +\frac{V_p}{2p}(re^{-\lambda}\dot H_p)\Bigr)\Bigr]\;.
\esea
Varying the Action~(\ref{Stau3}) w.r.t.\ $\lambda$ yields the
`reparametrization constraint' $C_1=0$ with
\bea
C_1&=&(\kappa-n)n-\frac{d-2p-1}{2p}(1-n^2)+\frac{1}{(r^2(1-n^2))^{p-1}}\cdot
\nonumber\\
&&\qquad\qquad\cdot\Bigl[
(d-2p-1)\Bigl(\frac{1}{2p}\frac{W^p}{r^2}+w^2H_p^2\Bigr)
-W^{p-1}u^2-\frac{V_p^2}{2p}\Bigr]\;.\qquad
\eea
Varying w.r.t.\ $n$, $\kappa$, $u$, and $V_p$ by construction yields
Eqs.~(\ref{deffirstford}) as field equations, and we will use them in the
remainig variations. Varying w.r.t.\ $\lambda$ for fixed $\nu+\lambda$
yields
\bsea\label{vareq}
re^{-\lambda}\dot n&=&(\kappa-n)n
-\frac{2}{(r^2(1-n^2))^{p-1}}
  \Bigl(W^{p-1}u^2+\frac{V_p^2}{2p}\Bigr)\;,
\\
\NoAlign{varying w.r.t.\ $r$ for fixed $e^{\nu+\lambda}\,r^{d-4p}$ yields}
re^{-\lambda}\dot\kappa&=&\Bigl(\frac{4p+1-d}{p}n-\kappa\Bigr)(\kappa-n)
\nonumber\\
&-&2(p-1)\frac{1-\kappa n}{1-n^2}\Bigl(
  (re^{-\lambda}\dot n)-\frac{d-2p-1}{2p}(1-n^2)\Bigr)
\nonumber\\
&+&\frac{2}{(r^2(1-n^2))^{p-1}}
  \Bigl(\frac{W^p}{r^2}-\frac{V_p^2}{2p}\Bigr)\;,
\\
\NoAlign{and varying w.r.t.\ $w$ and $H_p$ yields the remaining field
equations}
re^{-\lambda}\dot u&=&
(d-2p-1)\Bigl(\frac{w^2-1}{r}+r\frac{H_p^2}{W^{p-1}}\Bigr)w
\nonumber\\
&-&\Bigl(2(p-1)w\frac{ru}{w^2-1}+\kappa+(d-4p-1)n\Bigr)u\;,
\\
re^{-\lambda}\dot V_p&=&2p(d-2p-1)w^2H_p
  -\Bigl(\kappa+(d-4p)n\Bigr)V_p\;,
\esea

Next, we fix the freedom to reparametrise the radial variable $\tau$ by the
gauge choice $e^\lambda=r$ and define $V_p=(w^2-1)^{p-1}v$.

\subsection{New Variables -- Specific}
The equations of motion resulting from the action~(\ref{Stau3}) are singular when
$w^2=1$, due to the term~(\ref{YMH}) with $p>1$, or when $n^2=1$, due to the
term~(\ref{gp}) with $p>1$. In order to avoid these singularities, we have
to introduce new variables specifically adapted to the form of the
action~(\ref{Stau3}), whereas the procedure used to obtain this action is
essentially the same as used for other gravitating matter systems in
\cite{Breitenlohner:1994di, Breitenlohner:1993es, Breitenlohner:2003qj, Breitenlohner:2004fp,
Breitenlohner:2005hx}. The {\it conical fixed point\/} observed in
\cite{Breitenlohner:2005hx} is caused by the new variables required to
remove the singularity at $w^2=1$ from the equations of motion for that
particular model.

Here, we define new variables $\bu=u/t$, $\bh=h/t$, $\bv=v/t$ where
$t=(w^2-1)/r$, and introduce an additional, redundant variable
$y=t^2/(1-n^2)$, resulting in the system of differential equations with
polynomial r.h.s.\
\bsea\label{taueq}
\dot r&=&rn\;,
\qquad\qquad{\rm or}\quad
\dot s=-sn\;,
\qquad{\rm where}\quad
s=r^{-1}\;,
\\
\dot\nu&=&\kappa-n\;,
\\
\dot n&=&(\kappa-n)n
  -2y^p(1-n^2)\Bigl(\bu^2+\frac{\bv^2}{2p}\Bigr)\;,
\\
\dot\kappa&=&\Bigl(\frac{4p+1-d}{p}n-\kappa\Bigr)(\kappa-n)
+2y^p\Bigl[(1-n^2)\Bigl(\frac{d-2p-1}{2p}-\frac{\bv^2}{2p}\Bigr)
\nonumber\\&&\quad
+(p-1)(1-\kappa n)
    \Bigl((d-2p-1)(\frac{1}{2p}+w^2\bh^2)
         +\bu^2+\frac{\bv^2}{2p}\Bigr)\Bigr]\;,
\\
\dot y&=&\Bigl\{\Bigl(4w\bu+\frac{d-4p-1}{p}n\Big)
\nonumber\\&&\qquad
  -2y^p\,n\,\Bigl[(d-2p-1)\Bigl(\frac{1}{2p}+w^2\bh^2\Bigr)
  +\Bigl(\bu^2+\frac{\bv^2}{2p}\Bigr)\Bigr]\Bigr\}y\;,
\\
\dot w&=&(w^2-1)\bu\;,
\\
\dot\bu&=&(d-2p-1)\Bigl(1+(w^2-1)\bh^2\Bigr)w
  -\Bigl(2pw\bu+\kappa+(d-4p-2)n\Bigr)\bu\;,
\qquad
\\
\dot\bh&=&\bv-(2pw\bu-n)\bh\;,
\\
\dot\bv&=&2p(d-2p-1)w^2\bh
  -\Bigl(2pw\bu+\kappa+(d-4p-1)n\Bigr)\bv\;,
\esea
subject to the two constraints
\bsea\label{cons}
C_1&=&0\;,
\\
C_2&=&0\;,
\esea
where the expressions
\bsea\label{defcons}
C_1&=&(\kappa-n)n-(1-n^2)\Bigl[\frac{d-2p-1}{2p}
\nonumber\\&&\qquad
-y^p
  \Bigl((d-2p-1)(\frac{1}{2p}+w^2\bh^2)-\bu^2-\frac{\bv^2}{2p}\Bigr)\Bigr]\;,
\qquad\,
\\
C_2&=&r^2y(1-n^2)-(w^2-1)^2\;,
\esea
obey
\bsea\label{dcons}
\dot C_1&=&-2y^p\,n
  \Bigl(\frac{d-2p-1}{2}(1+2pw^2\bh^2)+(p-2)(\bu^2+\frac{\bv^2}{2p})\Bigr)C_1\;,
\\
\dot C_2&=&-r^2y\,nC_1+4w\bu C_2\;,
\esea
and thus the two constraints are preserved by, and therefore compatible with
the differential Eqs.~(\ref{taueq}).

Note that Eqs.~(\ref{taueq}b--i) are independent of $r$ (or $s$) and thus
remain regular for $r\to0$ as well as for $r\to\infty$, {\it i.e.}, for $s\to0$.

\subsection{Singular Points of the Differential Equations}
The differential Eqs.~(\ref{taueq}) determine the derivatives of the
dependent variables $Y=(r,Z)$ (or $Y=(s=r^{-1},Z)$) where $Z=(n, \kappa,
w, \bu, \bh, \bv, y)$ w.r.t.\ $\tau$,
\be
\dot Y=f(Y)\;.
\ee
They are singular when one of the dependent variables diverges and at the
fixed points (f.p.s) of the Dynamical System~(\ref{taueq}). For each such
f.p.\ $Y_0$ with $f(Y_0)=0$ we can introduce new variables $\tY=Y-Y_0$ and
linearise the equations
\be\label{dynsys}
\dot{\tY}=M\tY+O(\tY^2)\;.
\ee
Excluding the possibility that the matrix $M$ has eigenvalues with vanishing
real part, we can rewrite Eq.~(\ref{dynsys}) in terms of a suitable basis
$\tY=(\tY_-,\tY_+)$ as
\be
\frac{d}{d\tau}\left(\begin{array}{c}\tY-\\\tY+\end{array}\right)=
\left(\begin{array}{cc}-M_-&0\\0&M_+\end{array}\right)
\left(\begin{array}{c}\tY-\\\tY+\end{array}\right)+O(\tY^2)\;,
\ee
with positive definite matrices $M_-$ and $M_+$. Due to the theory of
dynamical systems there exists a `stable manifold'
$\tY=(\tY_-,\tY_+=O(\tY_-^2))$ of initial data such that $\tY\to0$ as
$\tau\to+\infty$ as well as an `unstable manifold'
$\tY=(\tY_-=O(\tY_+^2),\tY_+)$ such that $\tY\to0$ as $\tau\to-\infty$. 
Prop.~1 of~\cite{Breitenlohner:1993es} states conditions such that these
solutions can be characterised by and depend analytically on parameters
determined at the f.p.; this however is not possible in general.

First, there is the case $\bu\to\infty$, occuring when $w\to\pm1$ while
$\dot w=ru$ remains finite. Such solutions are of no interest since
Eqs.~(\ref{deffirstford}c) and~(\ref{vareq}c) exclude maxima of $|w|$ with
$|w|>1$.

Another type of singularity with $n\to0$ while $\kappa n\ne0$ occurs at
regular horizons and will be discussed below.

\subsubsection{Regular Origin}
A regular origin with $r\to0$ as $\tau\to-\infty$ is described by the f.p.\
with $n=\kappa=w=1$. Eq.~(\ref{taueq}g) then yields either
$\bu=-(d-2p-1)/2p$ or $\bu=1$, but only the second of these solutions is
compatible with a regular origin.

\subsubsubsection{Regular Origin without Higgs Field}
With $h\equiv0$ and $2p+1<d<4p+1$ Eq.~(\ref{taueq}e) implies
\be
y^p=1\;,
\ee
and suggests the Ansatz
\bsea\label{origw}
w(r)&=&1-b\,r^2+O(r^4)\;,
\\
u(r)&=&-2b\,r+O(r^3)\;,
\\
n(r)&=&1-c_n\,r^2+O(r^4)\;,
\\
\kappa(r)&=&1+c_\kappa\,r^2+O(r^4)\;,
\esea
where $b$ is a free parameter, whereas
\be\label{origwval}
c_n=c_\kappa=2b^2\;,
\ee
(compare \cite{Radu:2006mb}). Expressing Eqs.~(\ref{taueq}c--g) in terms of
the dependent variables $\tn=(1-n)/r^2$, $\tk=(\kappa-1)/r^2$,
$\tw=(w-1)/r^2$, $\tu=(\bu-1)/r$, and $\ty=(y-1)/r$ as functions of $r$
yields the equations
\bsea\label{origweq}
r\frac{d\tn}{dr}&=&\tn-\tk+rf_{\tn}\;,
\\
r\frac{d\tk}{dr}&=&(d-2)(\tn-\tk)+rf_{\tk}\;,
\\
r\frac{d\tw}{dr}&=&rf_{\tw}\;,
\\
r\frac{d\tu}{dr}&=&-d\,\tu+rf_{\tu}\;,
\\
r\frac{d\ty}{dr}&=&-d\,\ty+rf_{\ty}\;,
\\
\NoAlign{or, using the linear combinations $\tn_+=(d-2)\tn-\tk$ and
$\tn_-=\tn-\tk$,}
r\frac{d\tn_+}{dr}&=&r\Bigl((d-2)f_{\tn}-f_{\tk}\Bigr)\;,
\\
r\frac{d\tn_-}{dr}&=&-(d-3)\tn_-+r\Bigl(f_{\tn}-f_{\tk}\Bigr)\;.
\esea
The nonlinear terms $f_{\tn}$ etc.\ are analytic functions of $r$ and $\tn$
etc., and thus Eqs.~(\ref{origweq}c--g) fulfill the assumptions of Prop.~1
of~\cite{Breitenlohner:1993es}.  Furthermore Eqs.~(\ref{defcons}) can be
written as
\be
\frac{C_1}{r^2}=\tk-\tn+rf_{C_1}\;,
\qquad
\frac{C_2}{r^4}=2\tn-4\tw^2+rf_{C_2}\;,
\ee
and allow us to obtain the relation $\tn(0)=\tk(0)=2\tw^2(0)$. This
guarantees the local existence of a one parameter family of solutions with
the boundary Conditions~(\ref{origw}, \ref{origwval}), analytic in $r$ and
$b$, defined for all $b$ and $|r|<\xi(b)$ with some $\xi(b)>0$.

\subsubsubsection{Regular Origin with Higgs Field}
We supplement Eqs.~(\ref{origw}) with the Ansatz for the Higgs field,
\be\label{origh}
h(r)=a\,r+O(r^3)\;,
\qquad{\it i.e.,}\quad
\bh(r)=-\frac{a}{2b}+O(r^2)\;,
\ee
such that $\Phi(\vec x,t)$ is regular near $\vec x=0$. In view of
Eqs~(\ref{taueq}h,i) this requires $d=4p$ and consequently
\be\label{origv}
v(r)=(2p-1)a\,r+O(r^3)\;,
\qquad{\it i.e.,}\quad
\bv(r)=-(2p-1)\frac{a}{2b}+O(r^2)\;.
\ee
Evaluating Eq.~(\ref{taueq}e) at the f.p.\ yields
\be
y^p=\Bigl(1+(2p-1)\bh^2\Bigr)^{-1}\;.
\ee

We now have two free parameters, $a$ and $b$, whereas Eq.~(\ref{origwval})
is replaced by
\be\label{orighval}
c_n=2b^2(1+\gamma)^{1/p}\;,
\qquad
c_\kappa=\zeta(\gamma)c_n\;,
\qquad{\rm with}\quad
\zeta(\gamma)=1-\frac{2}{p}\,\frac{\gamma}{1+\gamma}\;,
\ee
and $\gamma=(2p-1)a^2/4b^2$. Expressing Eqs.~(\ref{taueq}c--i) in terms of
the dependent variables $\tn$, $\tk$, $\tw$, and $\tu$ as above together
with $\th_+=(2p\bh+\bv)/(4p-1)$, $\th_-=((2p-1)\bh-\bv)/((4p-1)r)$, and
$\ty=(y^p(1+\tg)-1)/r$ with $\tg=(2p-1)\th_+^2$ as functions of $r$ yields
the equations
\bsea\label{origheq}
r\frac{d\tn}{dr}&=&\zeta(\tg)\tn-\tk+rf_{\tn}\;,
\\
r\frac{d\tk}{dr}&=&(4p-2)\Bigl(\zeta(\tg)\tn-\tk\Bigr)+rf_{\tk}\;,
\\
r\frac{d\tw}{dr}&=&rf_{\tw}\;,
\\
r\frac{d\tu}{dr}&=&-4p\,\tu+rf_{\tu}\;,
\\
r\frac{d\th_+}{dr}&=&rf_{\th_+}\;,
\\
r\frac{d\th_-}{dr}&=&-4p\,\th_-+rf_{\th_-}\;,
\\
r\frac{d\ty}{dr}&=&-4p\,\ty+rf_{\ty}\;,
\\
\NoAlign{or, rewriting Eqs~(\ref{origheq}a--b) in terms of the new variables
$\tn_+=(4p-2)\tn-\tk$ and $\tn_-=\zeta(\tg)\tn-\tk$,}
r\frac{d\tn_+}{dr}&=&r\Bigl((4p-2)f_{\tn}-f_{\tk}\Bigr)\;,
\\
r\frac{d\tn_-}{dr}&=&-\Bigl(4p-2-\zeta(\tg)\Bigr)\tn_-
  +r\Bigl(\zeta(\tg)f_{\tn}-f_{\tk}
    -\frac{4}{p}\,\frac{2p-1}{(1+\tg)^2}\tn\th_+f_{\th_+}\Bigr)\;.
\qquad
\esea
Furthermore Eqs.~(\ref{defcons}) can be written as
\be
\frac{C_1}{r^2}=\tk-\zeta(\tg)\tn+rf_{C_1}\;,
\qquad
\frac{C_2}{r^4}=2\tn(1+\tg)^{-1/p}-4\tw^2+rf_{C_2}\;.
\ee
We use the Constraints~(\ref{cons}) to eliminate $\tn_-$ as well as
Eq.~(\ref{origheq}i) with the $\th_+$ dependent `eigenvalue'.  In addition
we obtain two relations between $\tn(0)$, $\tk(0)$, $\tw(0)$, and
$\th_+(0)$.  Eqs.~(\ref{origheq}c--h) then satisfy the assumptions of
Prop.~1 of~\cite{Breitenlohner:1993es}.  This guarantees the local existence
of a two parameter family of solutions with the boundary
Conditions~(\ref{origw}, \ref{origh}, \ref{origv}, \ref{orighval}), analytic
in $r$, $a$, and $b$, and defined for all $a$, $b\ne0$, and $|r|<\xi(a,b)$
with some $\xi(a,b)>0$.

\subsubsection{Regular Horizons}\label{subsectRH}
For regular, {\it i.e.}, non-degenerate horizons we may use $z=1/\kappa$ as the
independent variable tending to zero (compare~\cite{Breitenlohner:2004fp}), and thus
replace Eqs.~(\ref{taueq}) by
\bsea\label{hor}
z\frac{d}{dz}\tau&=&-\frac{\kappa}{\dot\kappa}=z(1+zf_\tau)\;,
\\
z\frac{d}{dz}y&=&z(1+zf_\tau)\dot y\;,
\\
z\frac{d}{dz}r&=&z(1+zf_\tau)\dot r\;,
\\
z\frac{d}{dz}w&=&z(1+zf_\tau)\dot w\;,
\\
z\frac{d}{dz}\bu&=&-\bu+zf_\bu\;,
\\
z\frac{d}{dz}\bh&=&z(1+zf_\tau)\dot\bh\;,
\\
z\frac{d}{dz}\bv&=&-\bv+zf_\bv\;,
\esea
with expressions $f_\tau$ etc.\ determined from Eqs.~(\ref{taueq}) that are
regular at $z=0$, while $n=O(z)$ can be computed from the
Constraint~(\ref{cons}a). These equations satisfy the assumptions of Prop.~1
of~\cite{Breitenlohner:1993es} and thus guarantee the existence of a family of solutions with
$\bu\to0$ and $\bv\to0$ as $z\to0$, and with finite limits $\tau_h$, $y_h$,
$r_h$, $w_h$, and $\bh_h$ for the remainig variables $\tau$, $y$, $r$, $w$,
and $\bh$. Finally, the Constraint~(\ref{cons}b) yields the relation
$y_h=(w_h^2-1)^2/r_h^2$.

In terms of the coordinate $\tau$ the behaviour near the horizon
is (performing a shift in $\tau$ such that $\tau_h=0$)
\bsea\label{bhbc}
r(\tau)&=&r_h\left(1+\frac{n_1\tau^2}{2}\right)+O(\tau^4)\;,
\\
n(\tau)&=&n_1\tau+O(\tau^3)\;,
\\
\kappa(\tau)&=&\tau^{-1}+O(\tau)\;,
\\
w(\tau)&=&w_h+\frac{r_hw_1\tau^2}{2}+O(\tau^4)\;,
\\
u(\tau)&=&w_1\tau+O(\tau^3)\;,
\\
h(\tau)&=&h_h\left(1-(p-1)\frac{r_hw_hw_1\tau^2}{w_h^2-1}\right)
  +\frac{h_1\tau^2}{2}+O(\tau^4)\;,
\\
v(\tau)&=&h_1\tau+O(\tau^3)\;,
\esea
with expressions $n_1$, $w_1$, and $h_1$ determined by the free parameters
$r_h$, $w_h$, and $h_h$.

We, thus, have a three parameter family of solutions that satisfy the black
hole boundary Conditions~(\ref{bhbc}) and are (except for the pole in
$\kappa$) analytic in $\tau$, $r_h$, $w_h$, and $h_h$, defined for all
$r_h$, $w_h$, $h_h$, and $|\tau|<\xi(r_h,w_h,h_h)$ with some
$\xi(r_h,w_h,h_h)>0$.

Solutions with a regular origin can be seen as the limits $r_h\to0$ of those with
a regular horizon. Consider a sequence of black hole solutions with finite
limits $h_h/r_h\to a_h$ and $(1-w_h)/r_h^2\to b_h$ as $r_h\to0$. Taken as
functions of $r$ they converge to a solution with regular origin and
parameters $a$ and $b$. Starting from the differential equations for $w(r)$ and
$h(r)$, derived form the action~(\ref{reducedactionGM}b), we can linearise
around $w=1$ and $h=0$, use a rescaled radial variable $\rho=r/r_h$, and
neglect all terms that vanish as $r_h\to0$, resulting in two hypergeometric
equations. We thus obtain the relations
\be\label{rhzero}
b^p={\frac{(p-1)d+p+1}{d-2p-1}\choose p-1}b_h^p\;,
\qquad{\rm and}\quad
a\,b^{p-1}={\frac{4p^2-3p+1}{2p-1}\choose p}a_hb_h^{p-1}\;,
\ee
valid for $d=4p$ or, without Higgs field, for $2p+1<d<4p+1$. Defining
the ratios $\zeta_b=b/b_h$ and $\zeta_a=a/a_h$, and evaluating the binomial
coefficients in Eqs.~(\ref{rhzero}) for $p\le4$ and $d=4p$ yields the values
shown in Tab.~\ref{tabzeta}, with the well known ratios $\zeta_b=1$ and
$\zeta_a=2$ for $p=1$ (compare Fig.~8b in \cite{Breitenlohner:1991aa}).
\begin{table}[ht]
\caption[tabzeta]{\label{tabzeta}Ratios between boundary conditions for
the limit $r_h\to0$ of black monopoles and the corresponding regular
monopoles for $p=1$, 2, 3, and~4 in $d=4p$ spacetime dimensions.}
\begin{center}
\begin{tabular}{l||l|l}
$p$& $\zeta_b$& $\zeta_a$
\\\hline\hline
1& 1& 2\\\hline
2& 1.91485422& 2.55313895\\\hline
3& 2.34407744& 2.81289292\\\hline
4& 2.60713329& 2.97958090\\\hline
\end{tabular}

\end{center}
\end{table}

\subsubsection{Reissner-Nordstr\"om Fixed Point}
The Reissner-Nordstr\"om (RN) f.p.\ is characterised by $n=w=u=v=0$,
$h=\alpha$, $r=1$, and $\kappa=\kappa_0$ where $\kappa_0^2=d-2p-1$, as for
the degenerate horizon of an extremal RN black hole. Introducing $\tr=r-1$,
$\tu=ru$, $\th=h-\alpha$, and $\tk=\kappa-\kappa_0$, we obtain
\bsea\label{linRN}
\dot\tr&=&n+\tr\,n\;,
\\
\dot w&=&\tu\;,
\\
\dot\tu&=&\kappa_0^2(\alpha^2-1)\,w-\kappa_0\,\tu+f_u\;,
\\
\dot\th&=&v+f_h\;,
\\
\dot v&=&-\kappa_0\,v+f_v\;,
\\
\dot n&=&\kappa_0\,n+f_n\;,
\\
\dot\tk&=&-2\kappa_0\,\tk+f_\kappa\;,
\esea
with expressions $f_i=O((\tr,w,\tu,\th,v,n,\tk)^2)$. Since $\tr$ can be
eliminated by the Constraint~(\ref{cons}a), and assuming $\kappa_0>0$, the
linearised Eqs.~(\ref{linRN}) have one unstable mode $n$ with eigenvalue
$\kappa_0$, one stable mode $\tk$ with eigenvalue $-2\kappa_0$. The
$(\th,v)$ subsystem contributes the stable mode $v$ with eigenvalue
$-\kappa_0$ and the zero mode $\th+v/\kappa_0$ (that has to be considered as
unstable due to the requirement $\th\to0$). The $(w,\tu)$ subsystem
contributes two modes with eigenvalues
\be\label{RNeig}
\lambda=-\frac{\kappa_0}{2}\Bigl(1\pm\sqrt{4\alpha^2-3}\Bigr)\;.
\ee
For $\alpha^2<3/4$ there are two stable oscillating modes with complex
conjugate eigenvalues with negative real part, for $3/4<\alpha^2<1$ there
are two stable modes with negative eigenvalues, whereas for $\alpha^2>1$
there is one stable and one unstable mode, {\it i.e.}, one positive and one
negative eigenvalue.

Thus, the dimension of the stable manifold is~4 for $\alpha^2<1$ and~3 for
$\alpha^2>1$, corresponding to a 3- resp.\ 2-parameter family of solutions
converging to the f.p.\ as $\tau\to+\infty$. However, it is possible to
extend the 3-dimensional stable manifold for $\alpha^2>1$ into the region
$3/4<\alpha^2\le1$ and define a 3-dimensional submanifold of the
4-dimensional stable manifold by the requirement that the variables in
Eqs.~(\ref{linRN}) decrease faster than $e^{-k_0\tau/2}$. When a family of
asymptotically flat solution reaches a critical limit with a double zero of
$\mu(r)$, the interior part with $r<1$ will be a member of the corresponding
2-parameter family, while the exterior part with $r>1$ will be the exterior
of the extremal RN solution with $w\equiv0$ and $h\equiv\alpha>\sqrt{3/4}$.

Note, however, that these solutions do not approach a degenerate horizon as
$\tau\to\infty$ and $r\to1$, because $e^\nu$ and hence $A=e^\nu/n$ diverge.
They describe geodesically complete spacetimes, asymptotically like
$AdS_2\times S^{d-2}$,
\be\label{AdSmetric}
ds^2\to e^{2\kappa_0\tau}dt^2-d\tau^2-d\Omega^2_{(d-2)}\;.
\ee

\subsubsection{Asymptotically Flat Infinity without Higgs Field}
Asymptotically flat infinity with $s\to0$ as $\tau\to+\infty$ is described
by a f.p.\ with $n=\kappa=1$. In the absence of a Higgs field this requires
$w=\pm1$ and Eq.~(\ref{taueq}g) again yields either
$\bu=-(d-2p-1)/2p$ or $\bu=1$, but this time only the first of these
solutions is compatible with asymptotically flat infinity.

Together with Eq.~(\ref{mass}), this suggests the Ansatz
\bsea\label{EYMasy}
n(s)&=&1-\frac{m}{2}z^2+O(z^4)\;,
\\
\kappa(s)&=&1+c_\kappa z^2+O(z^4)\;,
\\
w(s)&=&1-c\,z^2+O(z^4)\;,
\\
u(s)&=&\frac{d-2p-1}{p}c\,s\,z^2+O(sz^4)\;,
\esea
with $z=s^{(d-2p-1)/2p}$, where the total mass $M=\frac{4p+1-d}{2p}m^p$ and
$c$ are free parameters, while
\be
c_\kappa=\frac{(d-3p-1)m}{2p}\;.
\ee
Expressing Eqs.~(\ref{taueq}c--g) in terms of the dependent variables
$\tn=(1-n)/z^2$, $\tk=(\kappa-1)/z^2$, $\tw=(w-1)/z^2$,
$\tu=(\bu+(d-2p-1)/2p)/z$, and $\ty=y/(s\,z)^2$ as functions of $s$
yields the equations
\bsea\label{flatweq}
s\frac{d\tn}{ds}&=&\frac{3p+1-d}{p}\tn+\tk+z^2f_{\tn}\;,
\\
s\frac{d\tk}{ds}&=&\frac{d-3p-1}{p}\tn-\tk+z^2f_{\tk}\;,
\\
s\frac{d\tw}{ds}&=&z\,f_{\tw}\;,
\\
s\frac{d\tu}{ds}&=&-\Bigl(d-2+\frac{d-1}{2p}\Bigr)\tu+zf_{\tu}\;,
\\
s\frac{d\ty}{ds}&=&z^2f_{\ty}\;,
\\
\NoAlign{or, using the linear combinations $\tn_+=\tn+\tk$ and
$\tn_-=\tn(3p+1-d)/p+\tk$,}
s\frac{d\tn_+}{ds}&=&z^2\Bigl(f_{\tn}+f_{\tk}\Bigr)\;,
\\
s\frac{d\tn_-}{ds}&=&-\frac{d-2p-1}{p}\tn_-
  +z^2\Bigl(\frac{3p+1-d}{p}f_{\tn}+f_{\tk}\Bigr)\;.
\esea
The nonlinear terms $f_{\tn}$ etc.\ are analytic functions of $s$, $z$, and
$\tn$ etc., and thus, Eqs.~(\ref{flatweq}c--g) fulfill the assumptions of
Prop.~1 of~\cite{Breitenlohner:1993es} with $z$ as independent variable. Furthermore
Eqs.~(\ref{defcons}) can be written as
\be
\frac{C_1}{z^2}=\tn_-+z^2f_{C_1}\;,
\qquad
\frac{C_2}{z^4}=2\tn\ty-4\tw^2+z^2f_{C_2}\;,
\ee
and allow us to obtain the relations $\tk(0)=\tn(0)(d-3p-1)/p$ and
$\tn(0)\ty(0)=2\tw^2(0)$. This guarantees the local existence of a two
parameter family of solutions with boundary Conditions~(\ref{EYMasy}),
analytic in $M$, $c$, and $z$, defined for all $M$, $c$, and $|s|<\xi(M,c)$
with some $\xi(M,c)>0$.

\subsubsection{Asymptotically Flat Infinity with Higgs Field}
Asymptotically flat infinity with Higgs field is described by the f.p.\ with
$n=\kappa=1$, $w=u=v=0$, and $h=\alpha>0$.  The difficulty here is, that $w$
and $u$ decrease exponentially, approximately as $e^{-\sqrt{2p-1}\alpha r}$,
while $n$, $\kappa$, $h$, and $v$ converge much more slowly, as some power
of $s\equiv1/r$.

Assuming solutions with finite energy, we can use the
Definition~(\ref{mass}) and the Constraints~(\ref{defcons}) to obtain
expressions for $\tn=r(n-1)$ and $\tk=r(\kappa-1)$, that are bounded as long
as $w$, $u$, $h$, $v$, and the mass function $M(r)$ are bounded. This
suggests to derive differential equations for $w$, $u$, $h-\alpha$, $v$, and
$M$ with $r$ as independent variable
\bsea\label{EYMHasy}
\frac{dw}{dr}&=&u+f_w\;,
\\
\frac{du}{dr}&=&(2p-1)\alpha^2w+f_u\;,
\\
\NoAlign{with nonlinear terms $f_w$ and $f_u$ built from $w$, $u$,
$h-\alpha$, $v$, $\tn$, $\tk$, and $s$, while $d(h-\alpha)/dr$, $dv/dr$,
and}
\frac{dM}{dr}&=&(w^2-1)^{2(p-1)}\Bigl(\frac{2p-1}{2p}\frac{(w^2-1)^2}{r^2}
  +(2p-1)w^2h^2+u^2+\frac{v^2}{2p}\Bigr)\;,
\qquad
\esea
consist entirely of such nonlinear terms. This yields one dimensional stable
and unstable manifolds together with a `center' manifold. In order to
further analyze the orbits in the center manifold, in this particular case
simply $w=u=0$, we intoduce $\th=r(h-\alpha+v)$, $\tv=rv$ and obtain
\bsea\label{EYMHasycenter}
s\frac{dM}{ds}&=&s\,f_M\;,
\\
s\frac{d\th}{ds}&=&-\th+s\,f_\th\;,
\\
s\frac{d\tv}{ds}&=&s\,f_\tv\;,
\esea
with nonlinear terms $f_M$, $f_\th$, and $f_\tv$ that are bounded
expressions in terms of $M$, $\th$, and $\tv$. Consequently there exists a
three parameter family of solutions, partially characterised by the total
mass and the asymptotic value of $\tv$.

\section{Numerical Results}
\subsection{Numerical Procedure}
We are mainly interested in regular monopole solutions, {\it i.e.}, solutions of
Eqs.~(\ref{taueq}) connecting a regular origin, $r=0$, with asymptotically flat
infinity, $r\to\infty$. In addition there are `black monopoles' starting
from a non-degenerate horizon at $r=r_h$. To better understand some limiting
cases of these two types of solutions, we also need solutions ending at the
RN fixed point $w=0$, $h=\alpha$ at $r=1$ (joined with the exterior extremal
RN solution $w\equiv0$, $h\equiv\alpha$ for $r>1$).

We use a procedure that could be called `shooting and matching'.  For each
of the four cases mentioned, we use a suitable independent variable
$0\le\xi<\infty$. First we integrate Eqs.~(\ref{taueq}) from $\xi=0$ to some
$\xi=\xi_{\rm max}$ with a slightly modified Runge-Kutta algorithm, using
the shooting parameters $a$ and $b$ or $h_h$ and $w_h$ to replace
expressions that become undetermined at the singular starting point $\xi=0$.
Next we express the solution at $\xi=\xi_{\rm max}$ in terms of stable and
unstable modes suitable for the chosen endpoint $\xi\to\infty$. A solution
will only converge to that endpoint for data on the `stable manifold'
where (the values of) the unstable modes are functions of the stable modes
given by the solution of a system of integral equations. Finally, the
shooting parameters must be adjusted to satisfy the matching condition that
the solution at $\xi=\xi_{\rm max}$ lies on the stable manifold.

As usual for nonlinear systems, the procedure outlined above has a limited
domain of convergence and requires good approximate initial data. Since we
will study families of solutions depending smoothly on one or two
parameters, such approximate initial data are given by varying these
parameters in small steps.

In order to emphasise the similarities as well as the differences between
the members of the hierarchy of $p$-EBPS solutions, we present results
(figures and numbers) for $p=1$, 2, 3, and~4, repeating results for $p=1$
from \cite{Breitenlohner:1991aa, Breitenlohner:1994di} (for $\beta=0$, {\it i.e.},
without Higgs potential). In addition we will use results obtained in
\cite{Radu:2005rf} for the $p$-BPS hierarchy in flat space and in
\cite{Radu:2006mb} for the $p$-Bartnik-McKinnon (BK) hierarchy.

\subsection{Regular Monopoles}
As noted above, the limit $\alpha\to0$ can be taken in two different ways. 
The `unscaled' variables $\tilde r=\alpha r$ and $\tilde h=h/\alpha$ satisfy
$\alpha$-dependent equations together with the $\alpha$-independent boundary
condition $\tilde h\to1$ as $\tilde r\to\infty$.  For $\alpha=0$ this yields
the equations for the BPS monopole in flat space.  Near $\tilde r=0$ we have
$w=1-\tilde b\tilde r^2+O(\tilde r^4)$ and $\tilde h=\tilde a\tilde
r+O(\tilde r^3)$ with $\tilde a=\tilde b/\sqrt{2p-1}$ dictated by the
Bogomol'nyi equations, and $\tilde b=1/6$, $0.1623$, $0.13740$,
and~$0.11775$ for $p=1$, 2, 3, and~4 respectively (as found in
\cite{Radu:2005rf} but normalised differntly).  Starting from these values
we could increase $\alpha$ and adjust the parameters $\tilde a(\alpha)$ and
$\tilde b(\alpha)$ such that the boundary conditions $\tilde h\to1$ and
$w\to0$ as $\tilde r\to\infty$ remain satisfied.

Alternatively, we can use the `rescaled' variables $r$ and $h$ with the
$\alpha$-dependent boundary condition $h\to\alpha$ as $r\to\infty$.  Since
there is no Higgs potential, the resulting equations are
$\alpha$-independent.  For $\alpha=0$ and thus $h\equiv0$ they describe the
EYM system \cite{Radu:2006mb}.  Near $r=0$ we have $w=1-br^2+O(r^4)$ with
$b=\alpha^2\tilde b$ and $h=ar+O(r^3)$ with $a=\alpha^2\tilde a$, and thus
$a\approx b/\sqrt{2p-1}$ when $\alpha\ll1$.  In the absence of a Higgs
potential it suffices to adjust one of the parameters $a$ or $b$ such that
$w\to0$ as $r\to\infty$, whereas $\alpha$ is determined by the solution as
$\lim_{r\to\infty}h(r)$.

In addition to these fundamental monopole solutions with $w(r)\ge0$, there
exist excited solutions with $N=1$, 2,~$\dots$ zeros of $w(r)$. Starting
from the $N^{\rm th}$ generalised BK solutions with $a=0$ and $b=b_N^{(p)}$
as given in Tab.~\ref{tabBK}, we can increase $a$ and adjust $b(a)$ such
that $w(r)\to0$ as $r\to\infty$.

\begin{table}[ht]
\caption[tabBK]{\label{tabBK}Parameters of generalised Bartnik-McKinnon
solutions for $p=1$, 2, 3, and~4 in $d=4p$ spacetime dimensions.}
\begin{center}
\begin{tabular}{l||c|c|c|c||c}
$p$& $b_1^{(p)}$& $b_2^{(p)}$& $b_3^{(p)}$& $b_4^{(p)}$& $b_\infty^{(p)}$
\\\hline\hline
1& 0.453716& 0.651726& 0.697040& 0.704878& 0.706420\\\hline
2& 0.415609& 0.532402& 0.574678& 0.588191& 0.593799\\\hline
3& 0.410745& 0.502893& 0.539835& 0.554239& 0.562700\\\hline
4& 0.410127& 0.489756& 0.523142& 0.537311& 0.547598\\\hline
\end{tabular}
\end{center}
\end{table}

For $N\to\infty$ the BK solutions converge to a non-trivial limiting
solution for $r<1$, and to the exterior of an extremal RN black hole for
$r>1$. Starting from these limiting solutions with $a=0$ and
$b=b_\infty^{(p)}$ we obtain limiting monopole solutions.

The properties of the excited and limiting solutions are dominated by the
behaviour near the RN fixed point, where $w$ decreases exponentially (in
$\tau$) and oscillates with a frequency proportional to
$\sqrt{\alpha^2-0.75}$. Consequently all excited solutions coverge to the
limiting solution as $\alpha^2\to0.75$ and cease to exist beyond that
value.

\begin{figure}[ht]
\hbox to\linewidth{\hss
  \epsfig{bbllx=41bp,bblly=178bp,bburx=576bp,bbury=517bp,%
  file=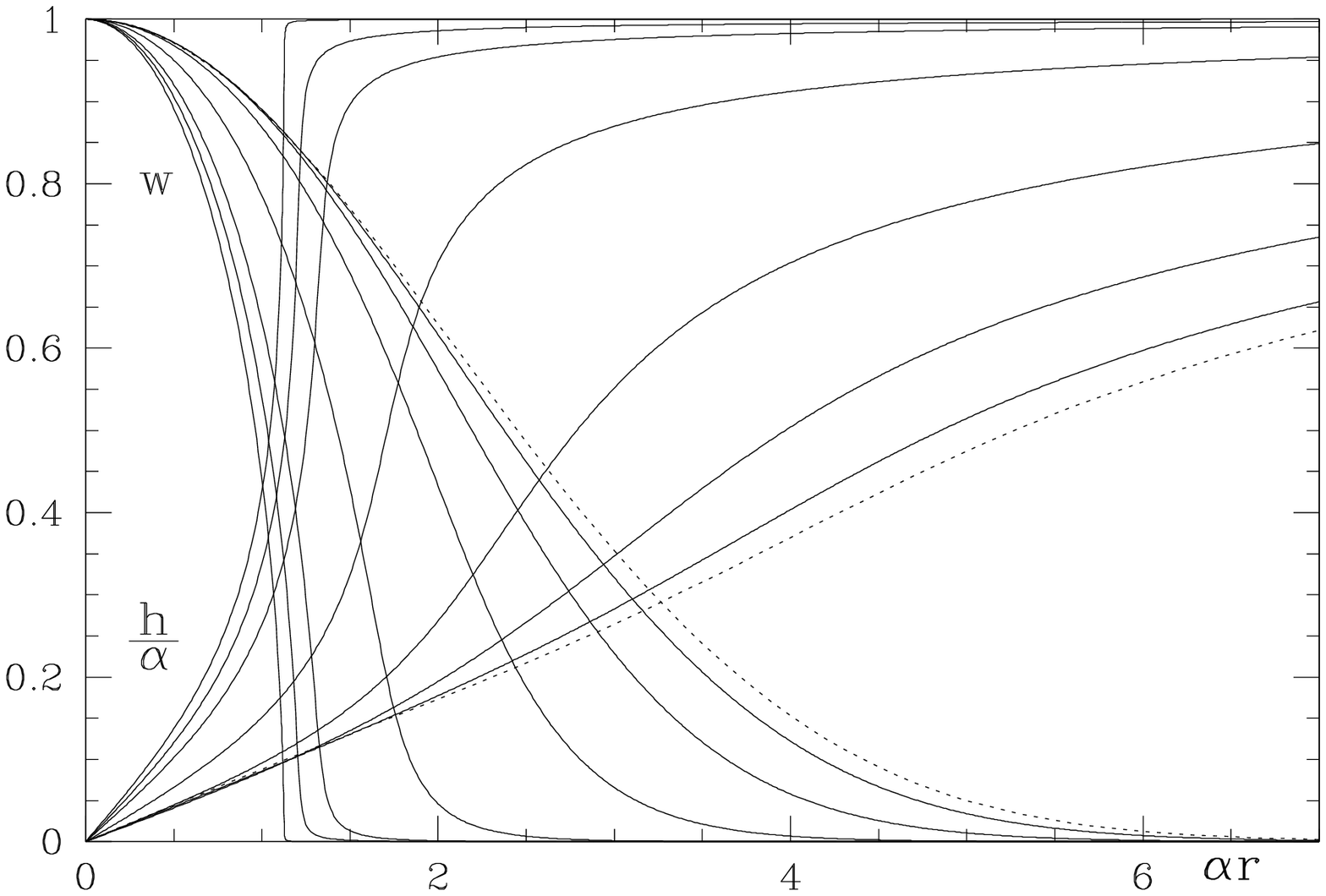,width=0.45\linewidth}\hss
  \epsfig{bbllx=41bp,bblly=178bp,bburx=576bp,bbury=517bp,%
  file=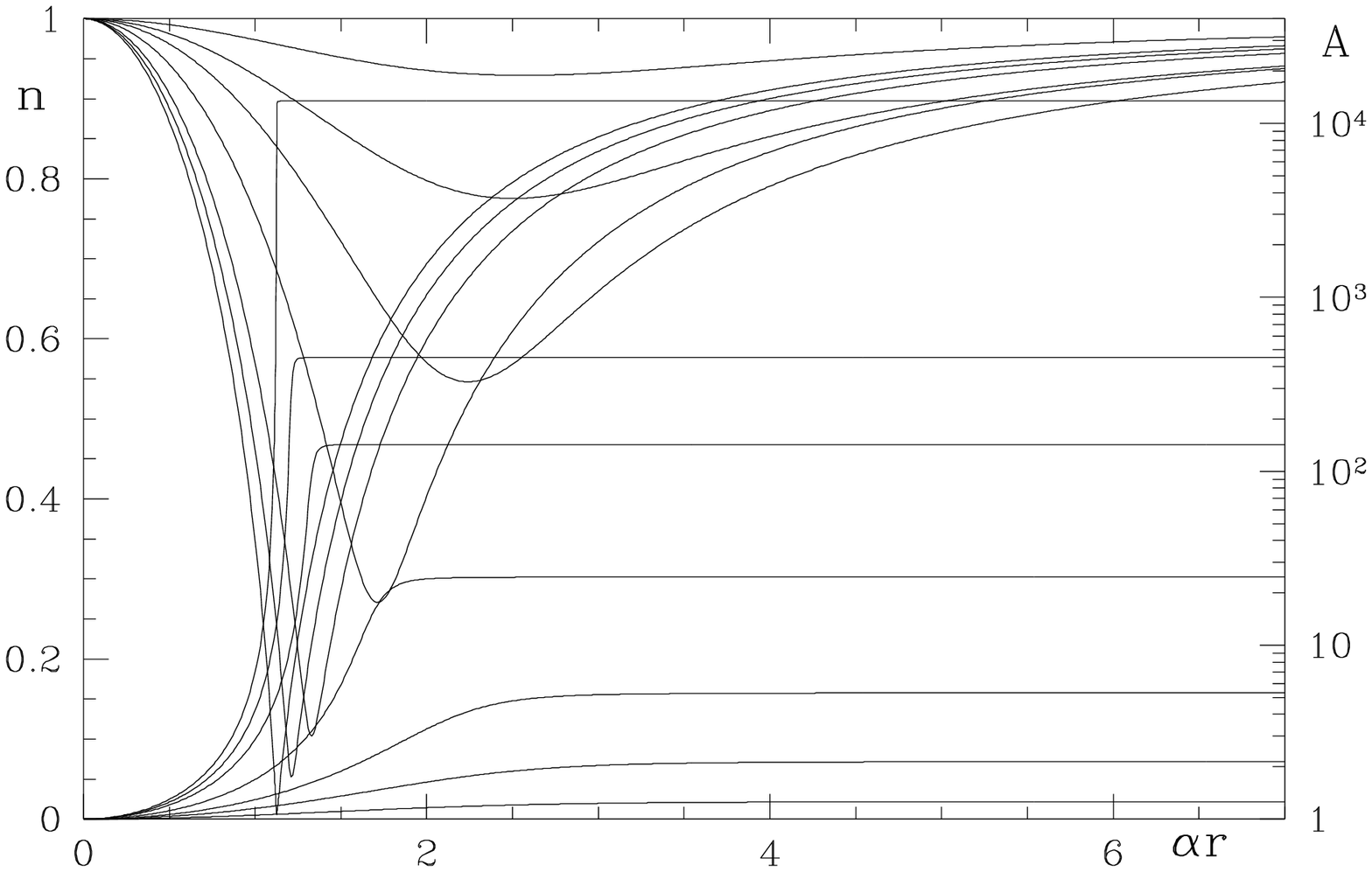,width=0.45\linewidth}\hss
}\vspace{-0.3cm}
\caption[figr0p4d]{\label{figr0p4d}Fundamental solutions
for $p=4$ with $\alpha=1.0$, $1.6$, $\alpha_{\rm max}$, $1.6$, $1.3$, $1.2$,
and $1.125$; the dotted curves are for the monopole in flat space.}
\end{figure}
\begin{figure}[ht]
\hbox to\linewidth{\hss
  \epsfig{bbllx=41bp,bblly=178bp,bburx=576bp,bbury=517bp,%
  file=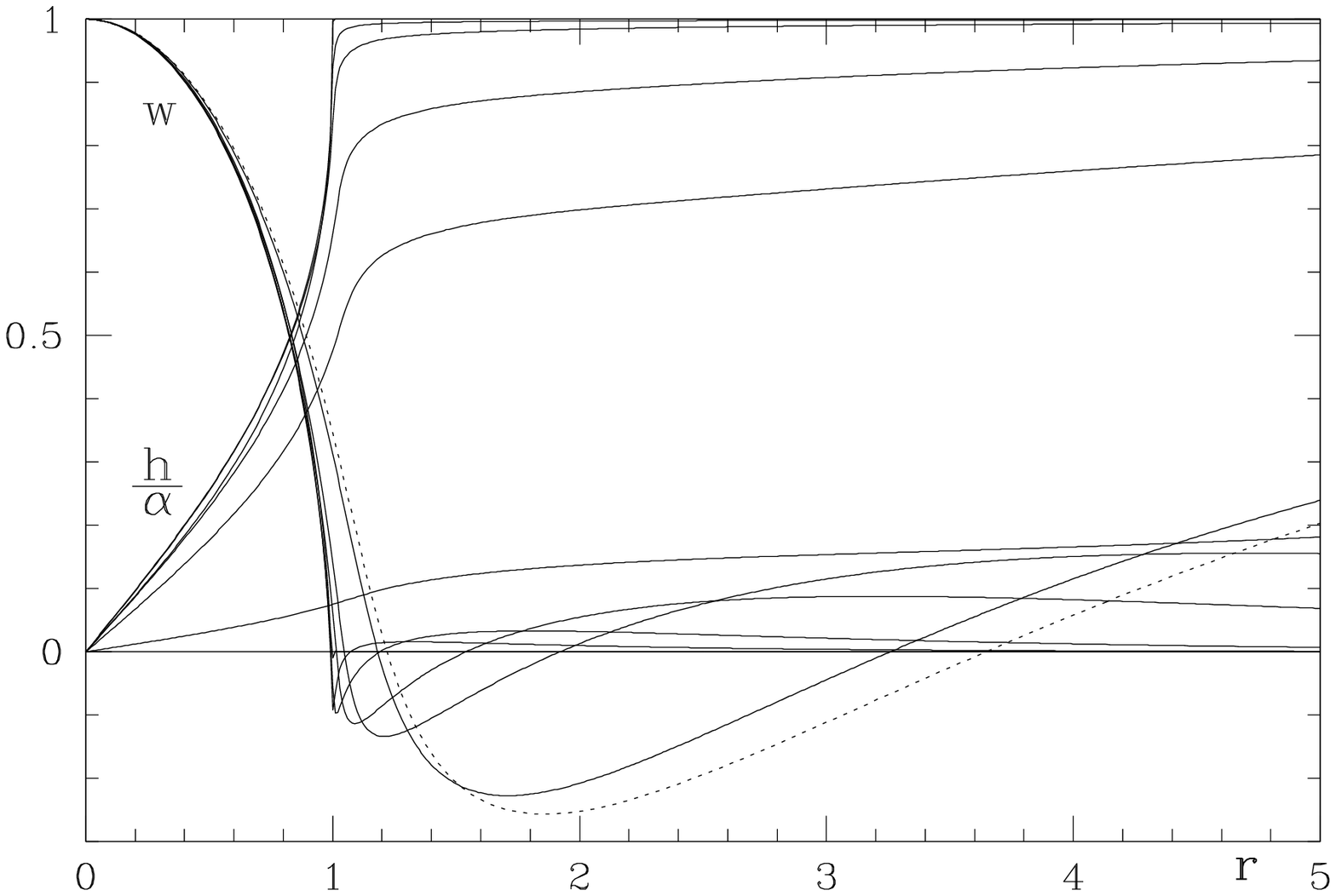,width=0.45\linewidth}\hss
  \epsfig{bbllx=41bp,bblly=178bp,bburx=576bp,bbury=517bp,%
  file=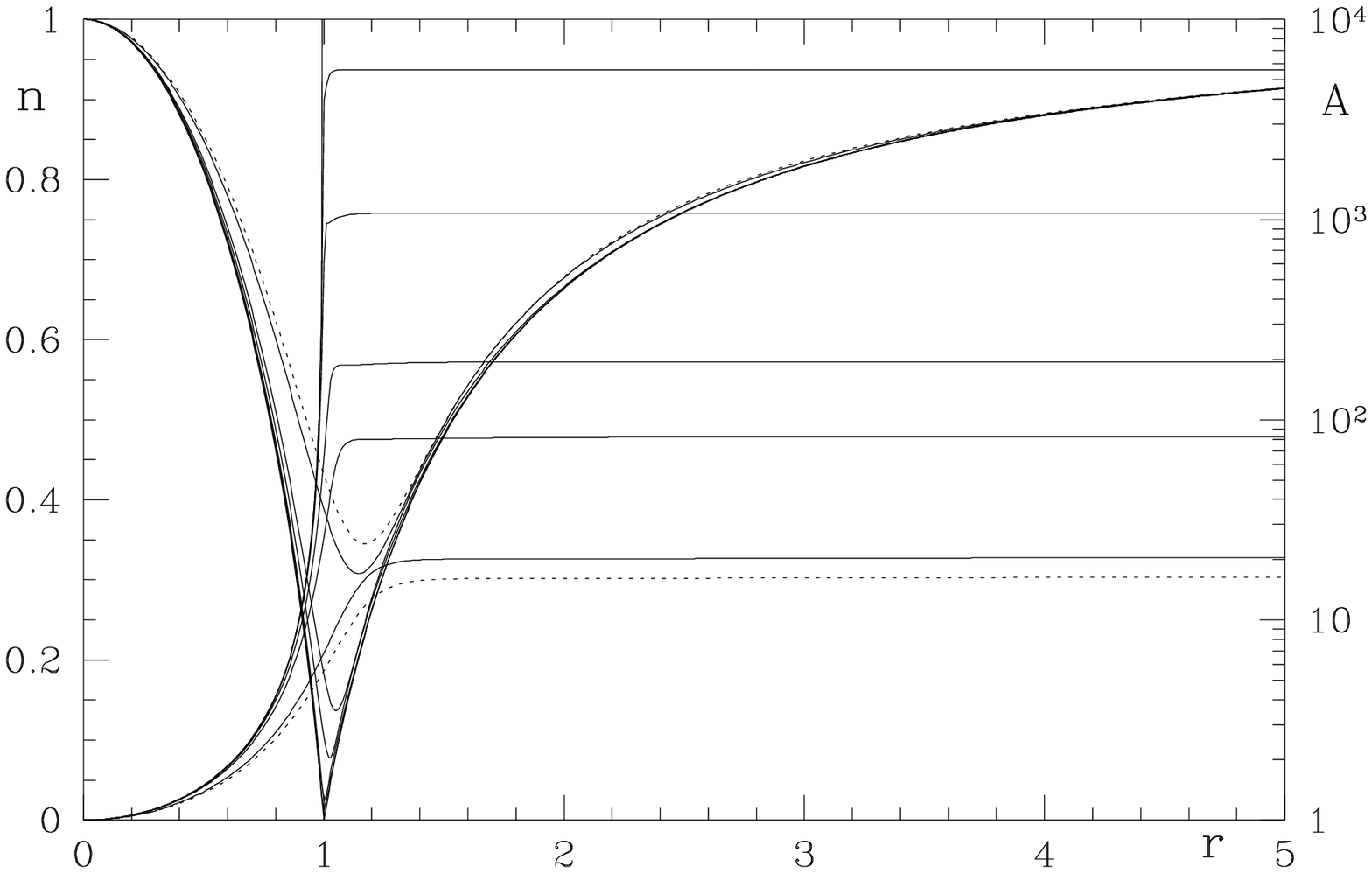,width=0.45\linewidth}\hss
}\vspace{-0.3cm}
\caption[figr2p2d]{\label{figr2p2d}Second excited solutions
for $p=2$ with $\alpha=0.1$, $0.15$, $0.2$, $0.35$, $0.5$ and $0.8$; the dotted
curves are for the corresponding BK solution.}
\end{figure}
\begin{figure}[ht]
\hbox to\linewidth{\hss
  \epsfig{bbllx=40bp,bblly=183bp,bburx=561bp,bbury=517bp,%
  file=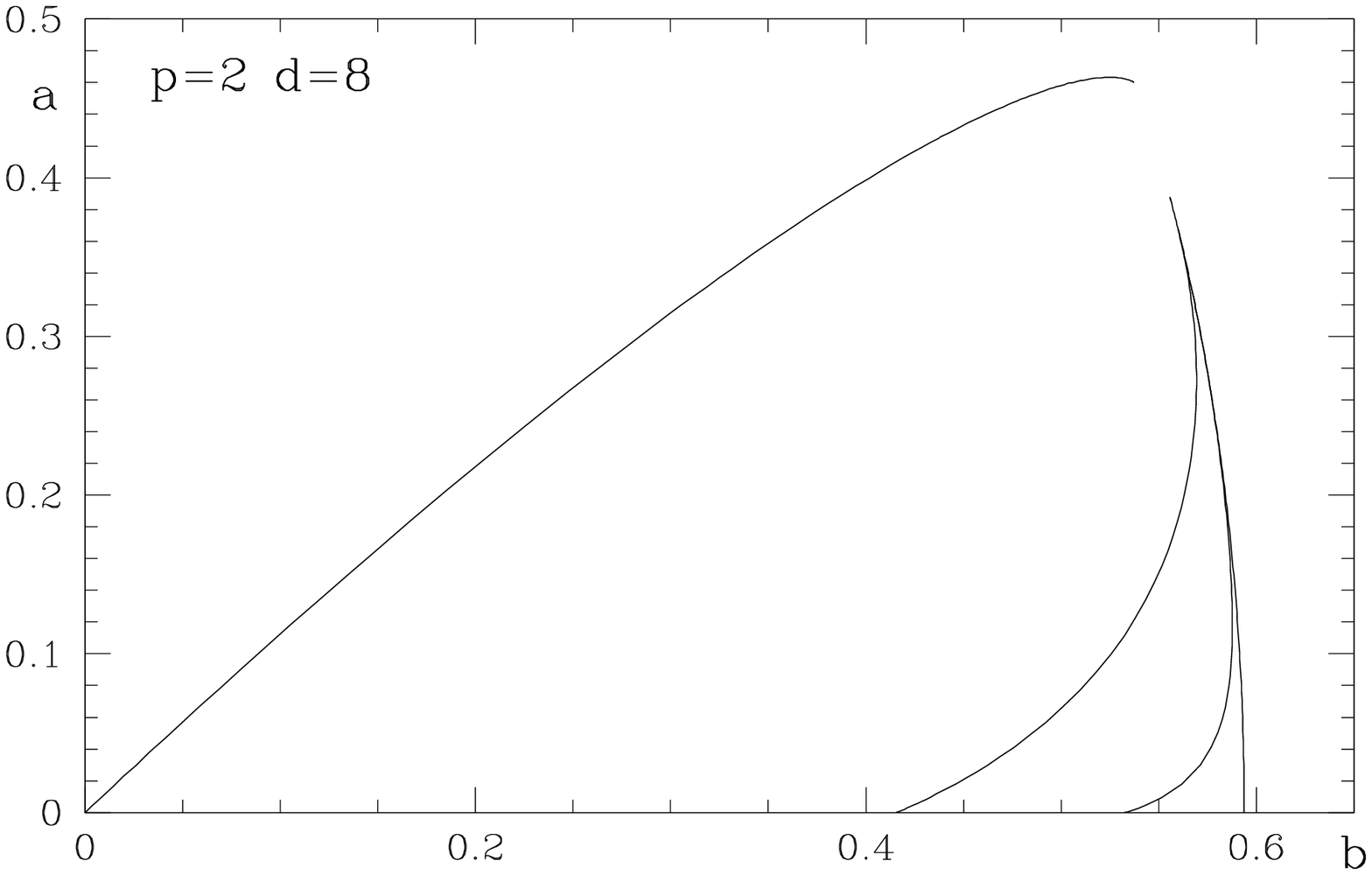,width=0.45\linewidth}\hss
  \epsfig{bbllx=40bp,bblly=183bp,bburx=561bp,bbury=517bp,%
  file=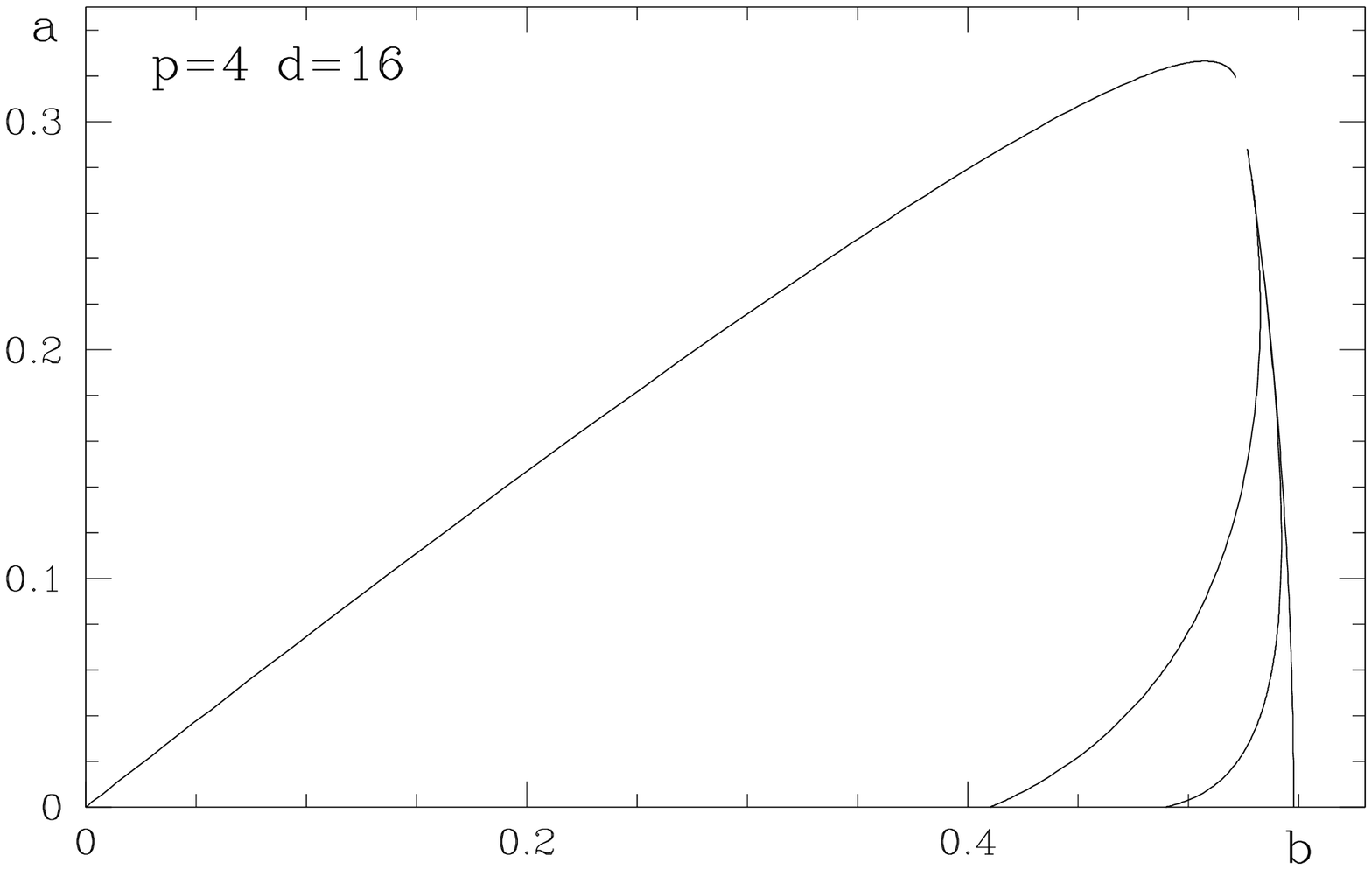,width=0.45\linewidth}\hss
}\vspace{-0.3cm}
\caption[figab]{\label{figab}Initial data for the regular monopole solutions
for $p=2$ and~4.}
\end{figure}

We thus obtain smooth one parameter families of solutions. Some fundamental
solutions for $p=4$ are shown in Fig.~\ref{figr0p4d} and $N=2$ excited ones
for $p=2$ are shown in Fig.~\ref{figr2p2d}. These families start with the
(rescaled) flat monopole for $N=0$ or the BK solutions for $N>0$ and end
with critical solutions where $\mu(r)$ has a double zero at $r=1$. These
critical solutions consist of two parts describing two geodesically complete
($t=$const) spaces: a non-trivial interior part for $r<1$ and a trivial part
with $w\equiv0$, $h\equiv\alpha$, and $A=$const for $r>1$, {\it i.e.}, the
exterior of an extremal RN black hole. The extremal RN black hole has a
degenerate horizon at $r=1$ and is a geodesically incomplete spacetime. The
non-trivial interior part describes a geodesically complete spacetime with
$A\to\infty$ as $r\to1$. The initial data for these families are quite
similar for $p=1$, 2, 3, and ~4 (see Fig.~\ref{figab}).

\begin{figure}[ht]
\hbox to\linewidth{\hss
  \epsfig{bbllx=40bp,bblly=183bp,bburx=561bp,bbury=517bp,%
  file=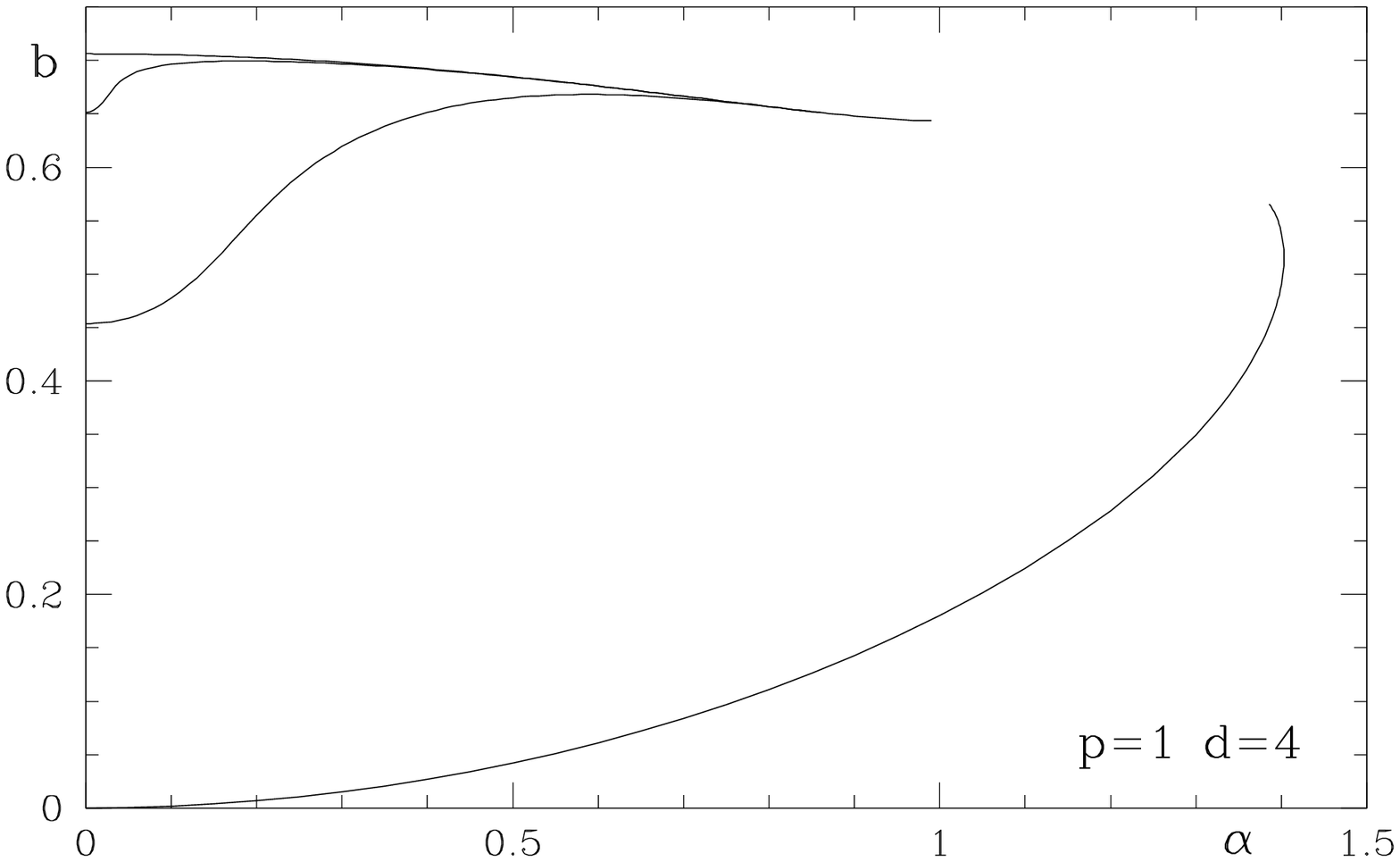,width=0.45\linewidth}\hss
  \epsfig{bbllx=40bp,bblly=183bp,bburx=561bp,bbury=517bp,%
  file=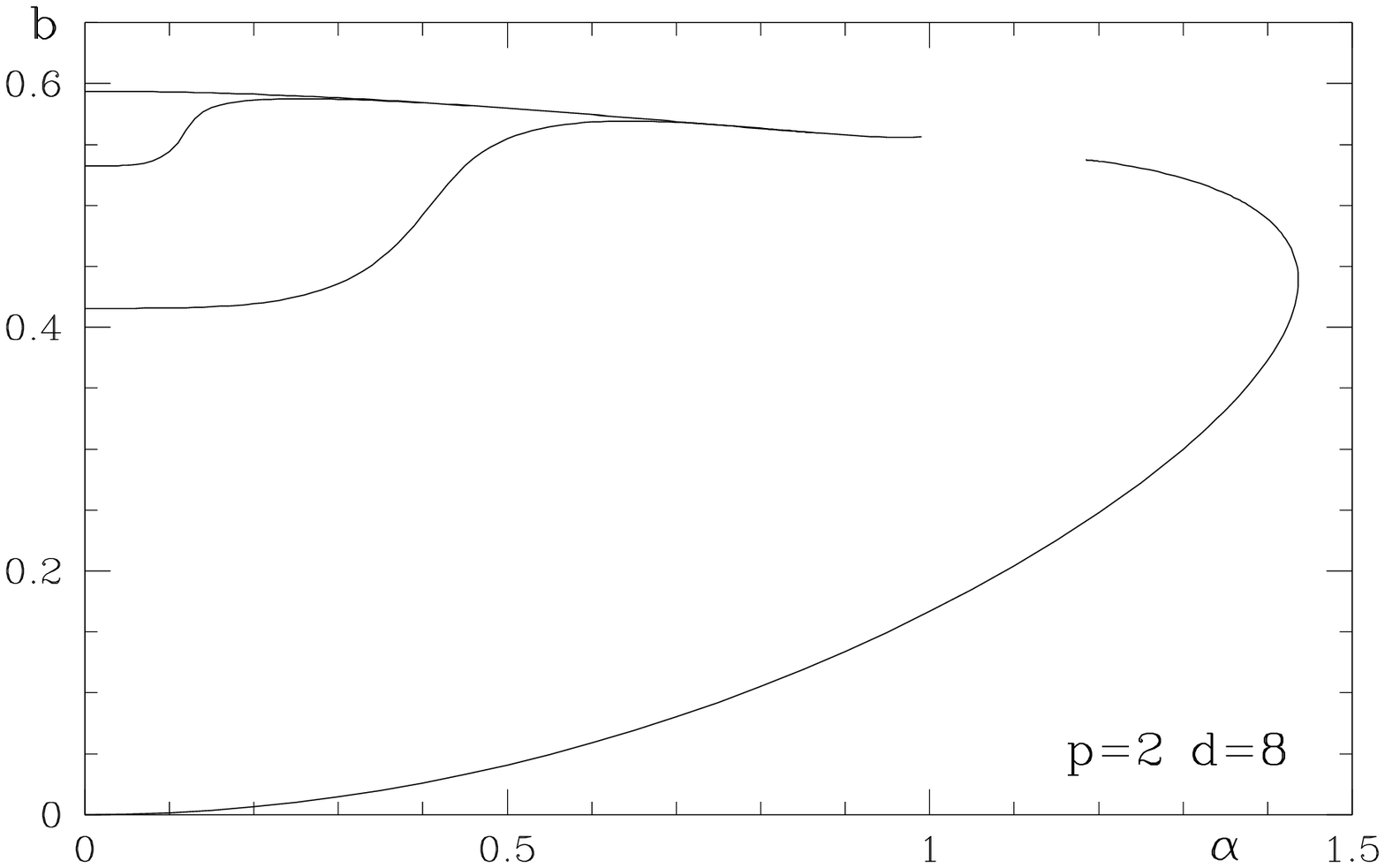,width=0.45\linewidth}\hss
}\vspace{0.2cm}
\hbox to\linewidth{\hss
  \epsfig{bbllx=40bp,bblly=183bp,bburx=561bp,bbury=517bp,%
  file=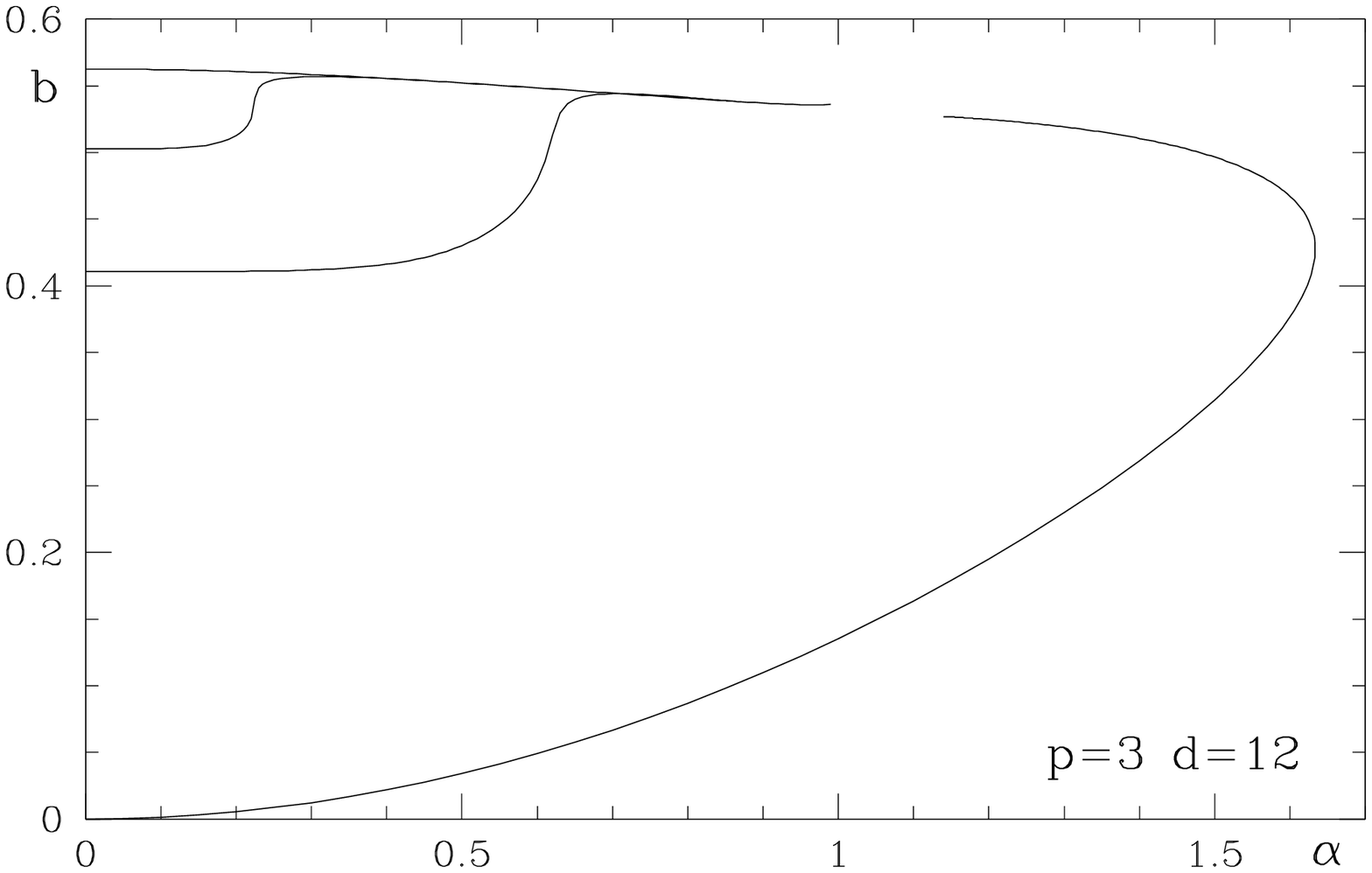,width=0.45\linewidth}\hss
  \epsfig{bbllx=40bp,bblly=183bp,bburx=561bp,bbury=517bp,%
  file=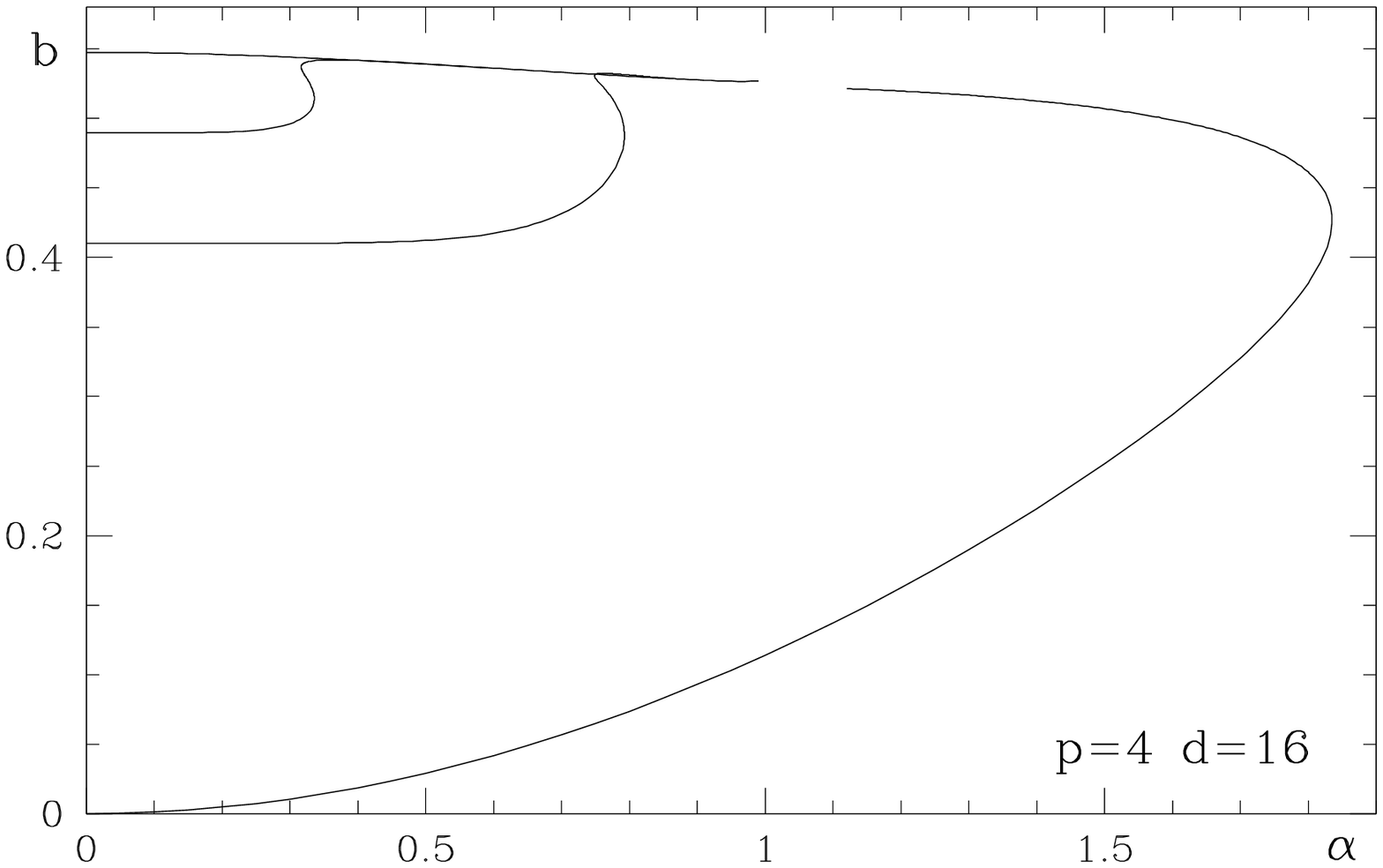,width=0.45\linewidth}\hss
}\vspace{-0.3cm}
\caption[figbalpha]{\label{figbalpha}$b$ {\it vs.}\ $\alpha$ for the regular
monopole solutions with $p=1$, 2, 3, and~4.}
\end{figure}

\begin{figure}[ht]
\hbox to\linewidth{\hss
  \epsfig{bbllx=40bp,bblly=183bp,bburx=561bp,bbury=517bp,%
  file=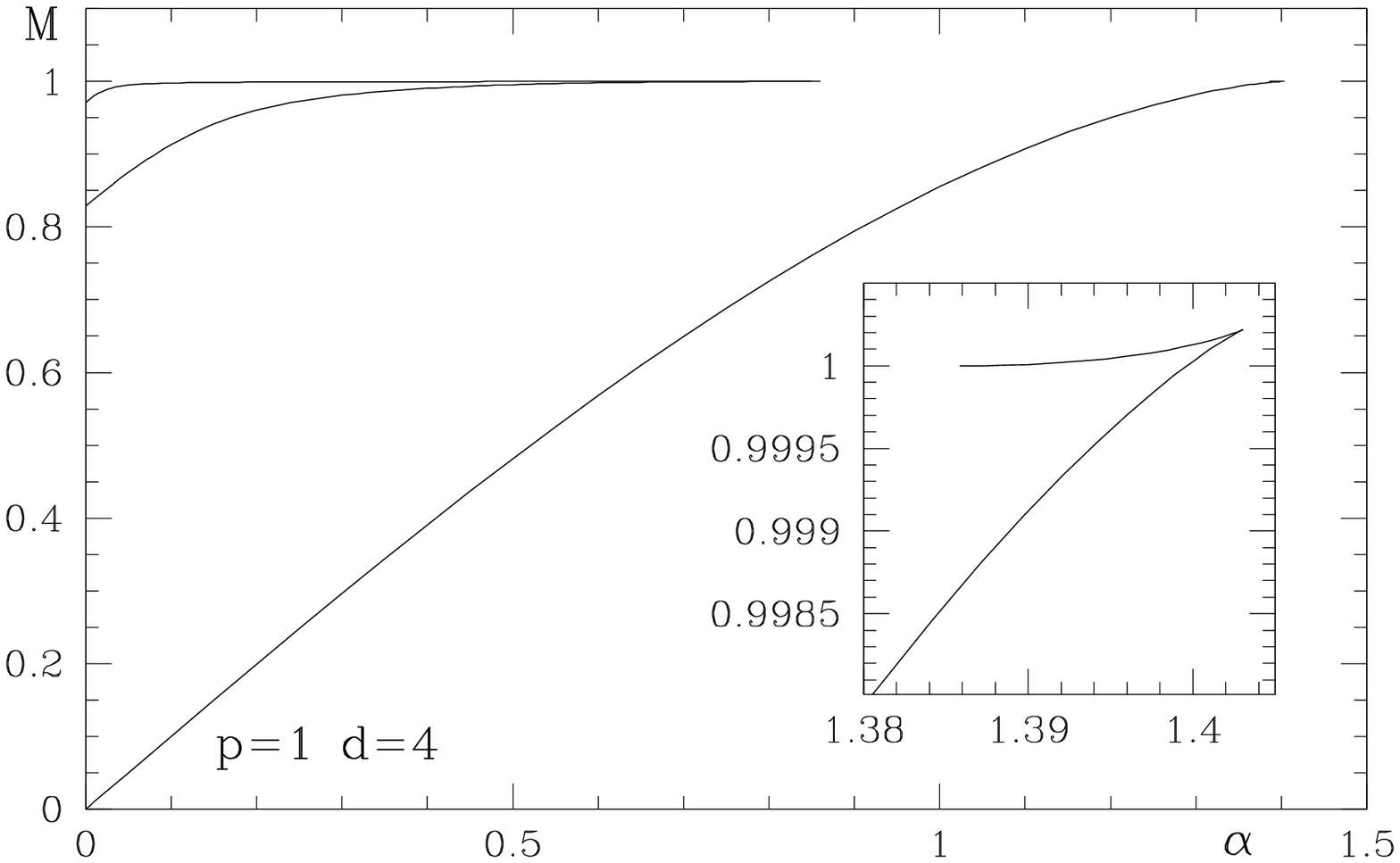,width=0.45\linewidth}\hss
  \epsfig{bbllx=40bp,bblly=183bp,bburx=561bp,bbury=517bp,%
  file=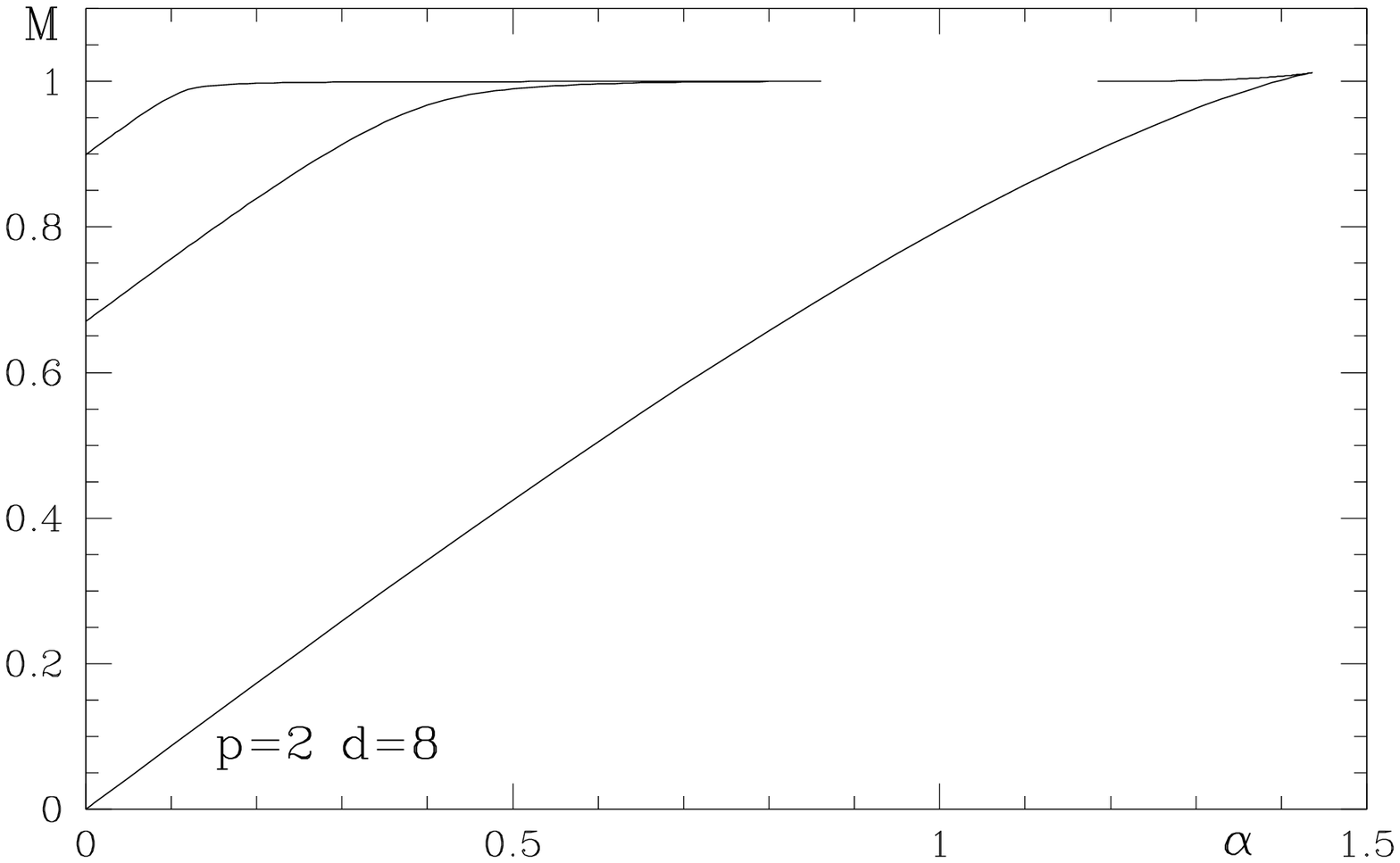,width=0.45\linewidth}\hss
}\vspace{0.2cm}
\hbox to\linewidth{\hss
  \epsfig{bbllx=40bp,bblly=183bp,bburx=561bp,bbury=517bp,%
  file=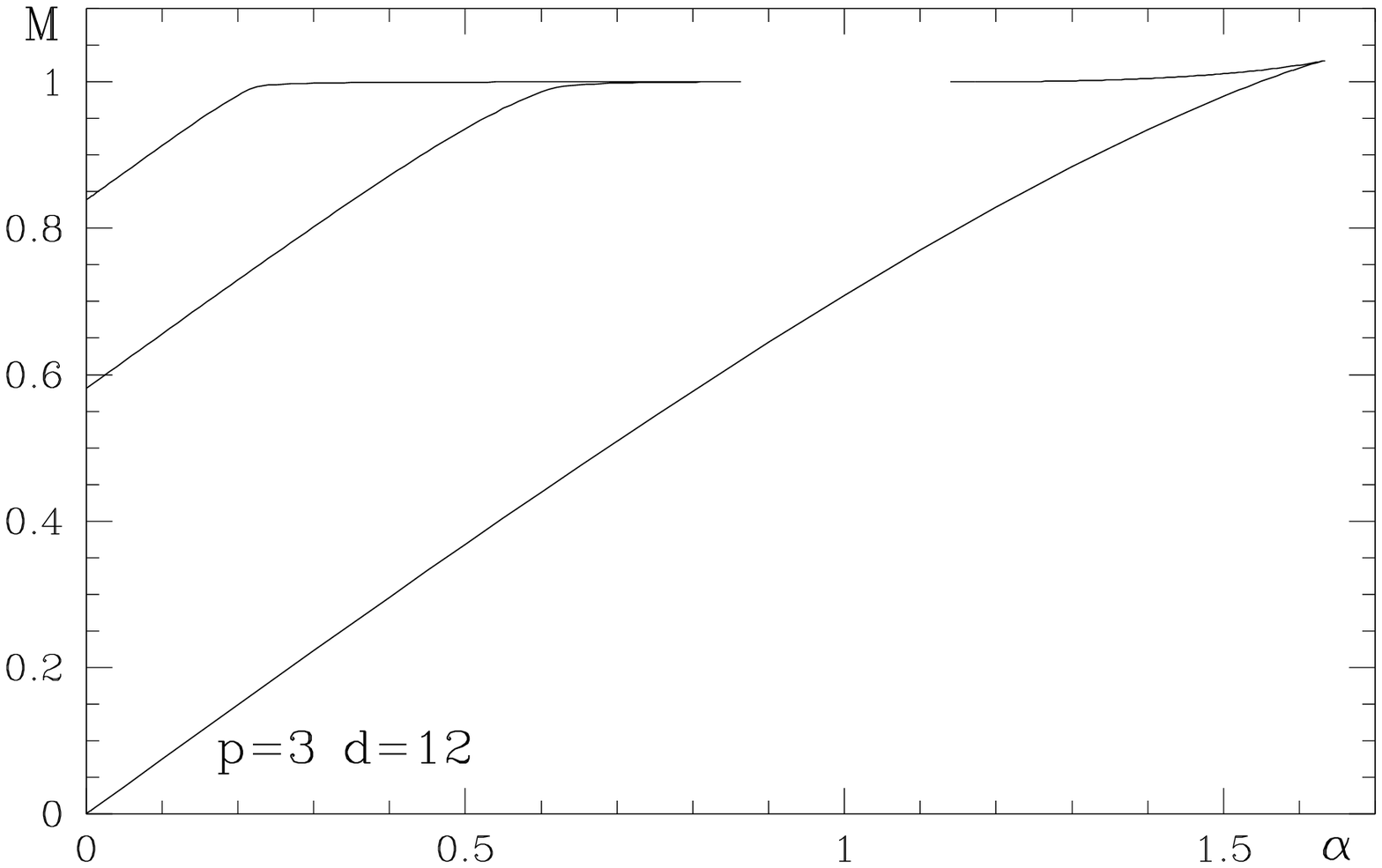,width=0.45\linewidth}\hss
  \epsfig{bbllx=40bp,bblly=183bp,bburx=561bp,bbury=517bp,%
  file=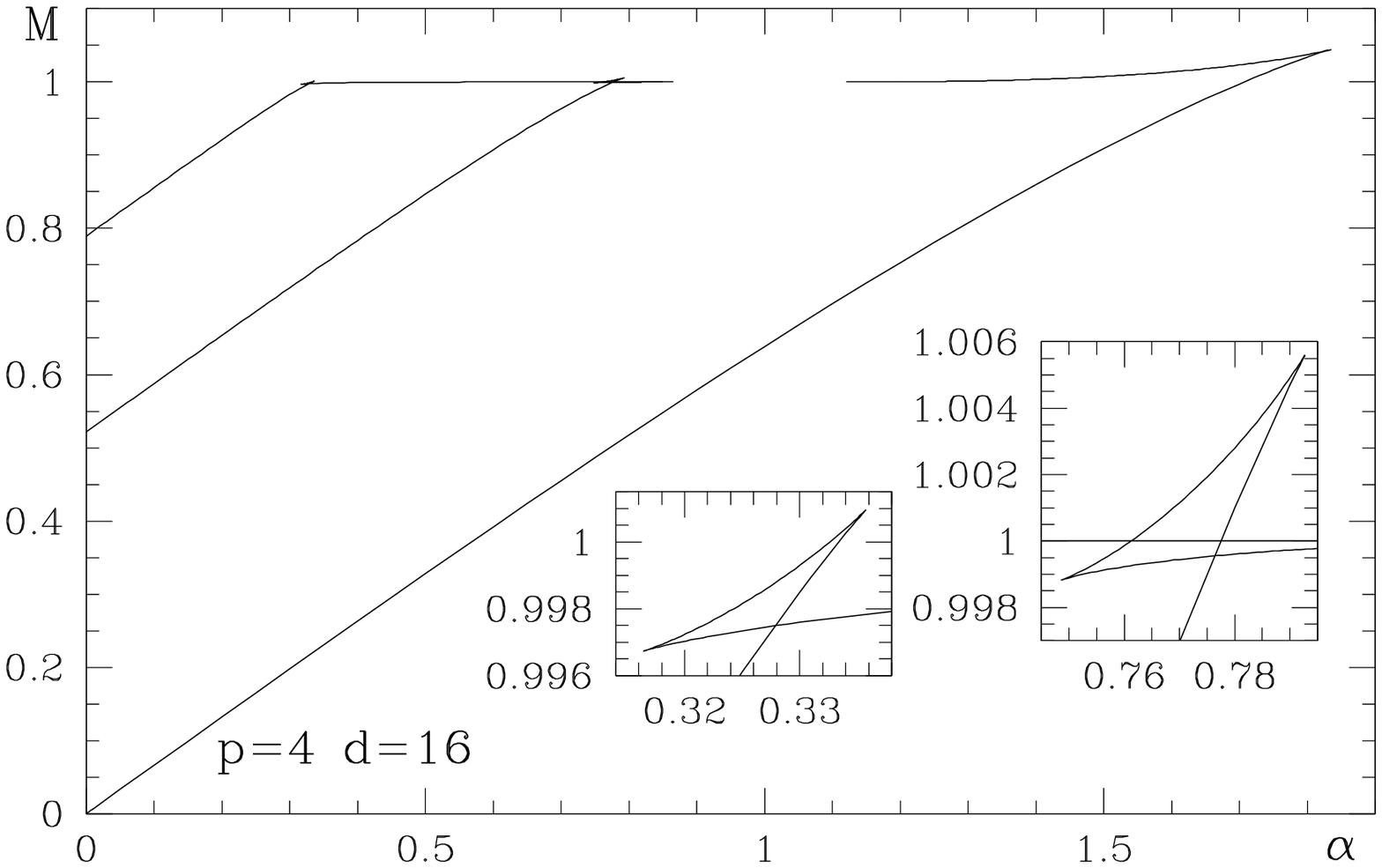,width=0.45\linewidth}\hss
}\vspace{-0.3cm}
\caption[figmalpha]{\label{figmalpha}Mass of the regular monopole solutions
with $p=1$, 2, 3, and~4.}
\end{figure}

An interesting structure emerges when we include the values of $\alpha$
obtained from these solutions. For $N=0$ and $1\le p\le4$, $\alpha$ first
increases from 0 to some $\alpha_{\rm max}$ and subsequently decreases to
some $\alpha_{\rm cr}$. Simultaneously the mass $M$ starts as $\alpha$ times
the mass $M_{\rm flat}=\sqrt{2p-1}/p$ of the flat space monopole, increases
to a maximum $M_{\rm max}>1$ and subsequently decreases to $M_{\rm cr}=1$
(see Figs.~\ref{figbalpha} and~\ref{figmalpha}). In Tab.~\ref{tabMalpha} we
have collected some relevant numbers. The difference $\alpha_{\rm
max}-\alpha_{\rm cr}$ is very small for $p=1$, but quite large for $p>1$.

\begin{table}[ht]
\caption[tabMalpha]{\label{tabMalpha}Maximal and critical values of $\alpha$
and masses of the fundamental monopole solutions for $p=1$, 2, 3, and~4.}
\begin{center}
\begin{tabular}{l||c|c|c}
$p$& $\alpha_{\rm max}$& $M_{\rm max}$& $\alpha_{\rm cr}$
\\\hline\hline
1& 1.40303& 1.00022& 1.38585\\\hline
2& 1.43629& 1.01212& 1.18516\\\hline
3& 1.63363& 1.02921& 1.14018\\\hline
4& 1.83513& 1.04432& 1.12034\\\hline
\end{tabular}
\end{center}
\end{table}

This may be a good point to comment on the numerical procedure described
above. The possible values of $\xi_{\rm max}$ are limited severely by the
necessity to suppress the unstable YM mode growing exponentially with
$\xi\equiv r$, whereas $h(r)$ converges only slowly to its asymptotic value
$h(\infty)=\alpha$. For the $p=4$ solution with $\alpha_{\rm max}$ as shown
in Tab.~\ref{tabMalpha}, $h(r)$ differs by more than 10\,\% for the largest
possible values of $\xi_{\rm max}$. Requiring that the unstable Higgs mode,
$\th$ in Eq.(\ref{EYMHasycenter}b) vanishes at $\xi_{\rm max}$ yields
excellent results for the BPS monopole in flat space. But in the gravitating
case considered here, $f_{\th}$ has slowly decreasing contributions
proportional to $\tn$ or $\tk$, preventing to estimate $h(\infty)$ with an
error much better than $10^{-3}$. The results obtained with `shooting and
matching' as outlined, however, are accurate up to rounding and
discretization errors, typically around $10^{-10}$ and are with that
accuracy independent of the choice of $\xi_{\rm max}$.

The (first three) excited solutions for $p=1$, 2, and~3 show a monotonic
increase of $\alpha$ from 0 to $\alpha_{\rm cr}=\sqrt{0.75}$, with the mass
increasing from its BK value to~1. For $p=4$, however, the values of
$\alpha$ first increase to some $\alpha_{\rm max}$ with $M_{\rm max}>1$,
then decrease to some $\alpha_{\rm min}$ with $M_{\rm min}<1$, and finally
increase to $\alpha_{\rm cr}$ with $M=1$ (see Figs.~\ref{figbalpha}
and~\ref{figmalpha}). One might speculate that higher excited solutions,
with $N>3$, exhibit such maxima and minima also for $p<4$.

The maxima of $\alpha$ (and $M$) observed for the fundamental monopoles give
rise to two `branches' of solutions. Furthermore, for $\alpha=\alpha_{\rm
max}$ there exists a `zero mode', indicating a change in the number of
instabilities against small time-dependent perturbations. Whereas the
`lower' branch from 0 to $\alpha_{\rm max}$ inherits the stability of the
flat space BPS monopole, the `upper' branch from $\alpha_{\rm max}$ to
$\alpha_{\rm cr}$ has one such instability.

For $\alpha\ll1$ the excited monopoles with mass $\approx M_{\rm BK}+\alpha
M_{\rm flat}$ may be seen as `superposition' of a BK solution for $\alpha
r\ll1$ and a flat space monopole for $r\gg1$, and inherit the instabilities
of the BK solution. The $N^{\rm th}$ BK solution for $p=1$ has $2N$ such
instabilities ($N$ radial and $N$ sphaleronic ones) and this might equally
be true for $p>1$. The maxima and minima of $\alpha$ (and $M$) observed for
the excited $p=4$ monopoles give rise to three branches of solutions,
where the `middle' branch from $\alpha_{\rm max}$ to $\alpha_{\rm min}$ has
one additional instability.

\subsection{Black Monopoles}
As noted in Subsect.~\ref{subsectRH} the limits $r_h\to0$ of black monopole
solutions yields regular monopoles. Conversely the (fundamental, excited, or
limiting) regular monopoles can be taken as starting points for corresponding
black monopoles with $r_h\ll1$.

Using `rescaled' variables one can, {\it e.g.}, choose a value $r_h$ and adjust
one of the parameters $h_h$ or $w_h$ such that $w\to0$ as $r\to\infty$,
whereas $\alpha$ is again determined by the solution as
$\lim_{r\to\infty}h(r)$. We thus obtain smooth two parameter families of
solutions.

\begin{figure}[ht]
\hbox to\linewidth{\hss
  \epsfig{bbllx=40bp,bblly=183bp,bburx=561bp,bbury=517bp,%
  file=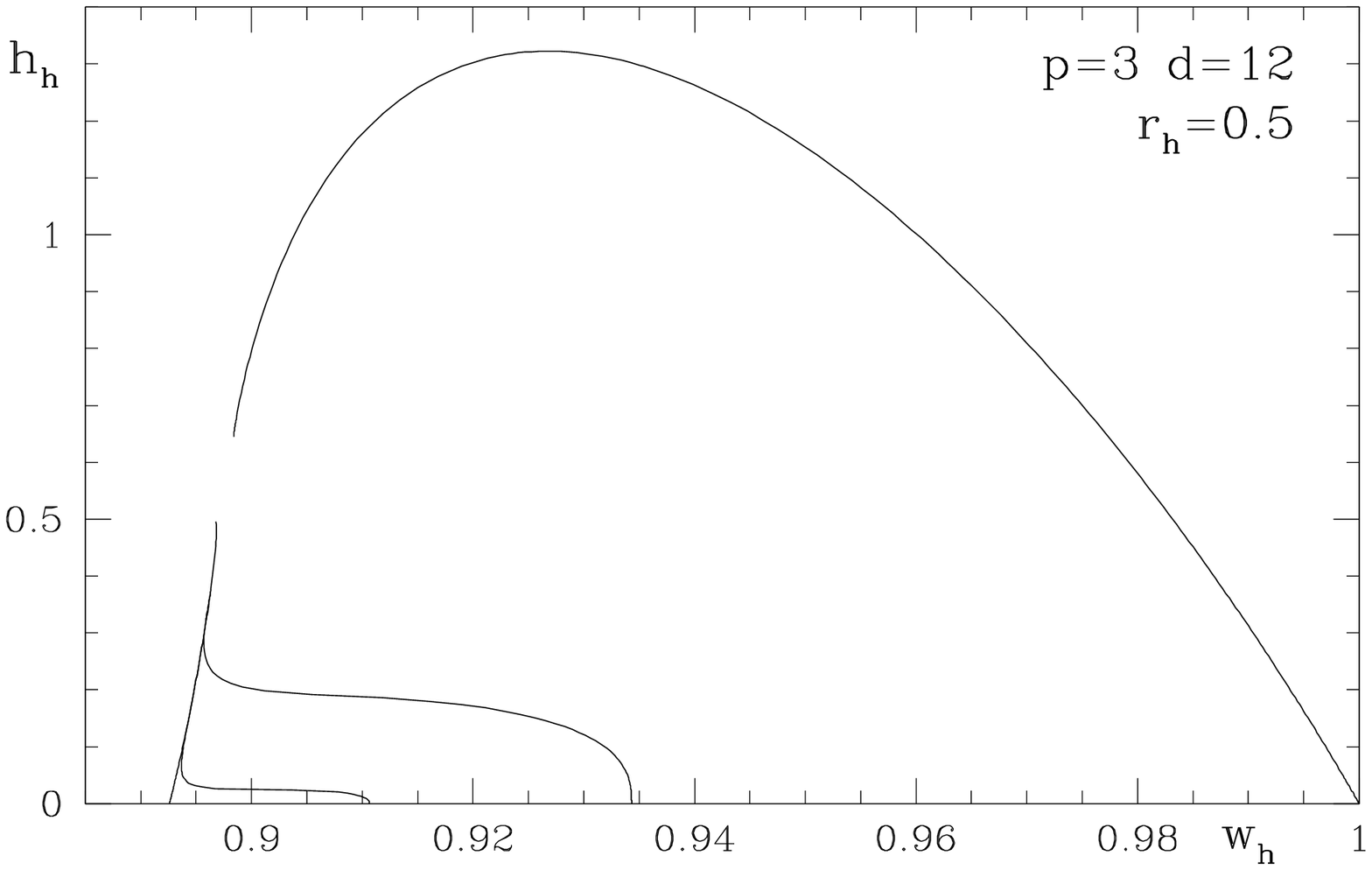,width=0.45\linewidth}\hss
  \epsfig{bbllx=40bp,bblly=183bp,bburx=561bp,bbury=517bp,%
  file=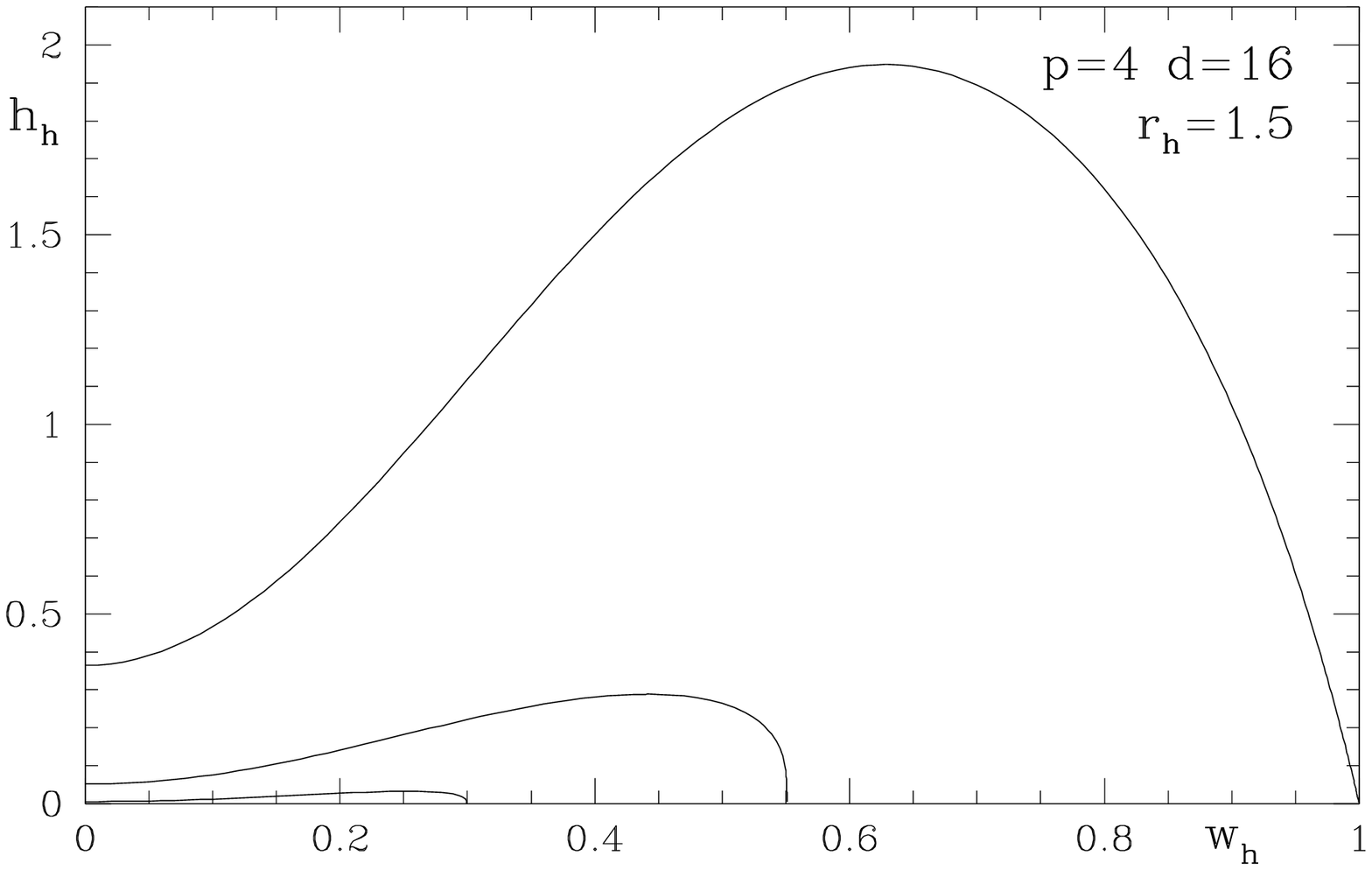,width=0.45\linewidth}\hss
}\vspace{-0.3cm}
\caption[figphwh]{\label{figphwh}Initial data for the black monopole
solutions with $r_h=0.5$ for $p=3$ and $r_h=1.5$ for $p=4$.}
\end{figure}
\begin{figure}[ht]
\hbox to\linewidth{\hss
  \epsfig{bbllx=41bp,bblly=178bp,bburx=576bp,bbury=517bp,%
  file=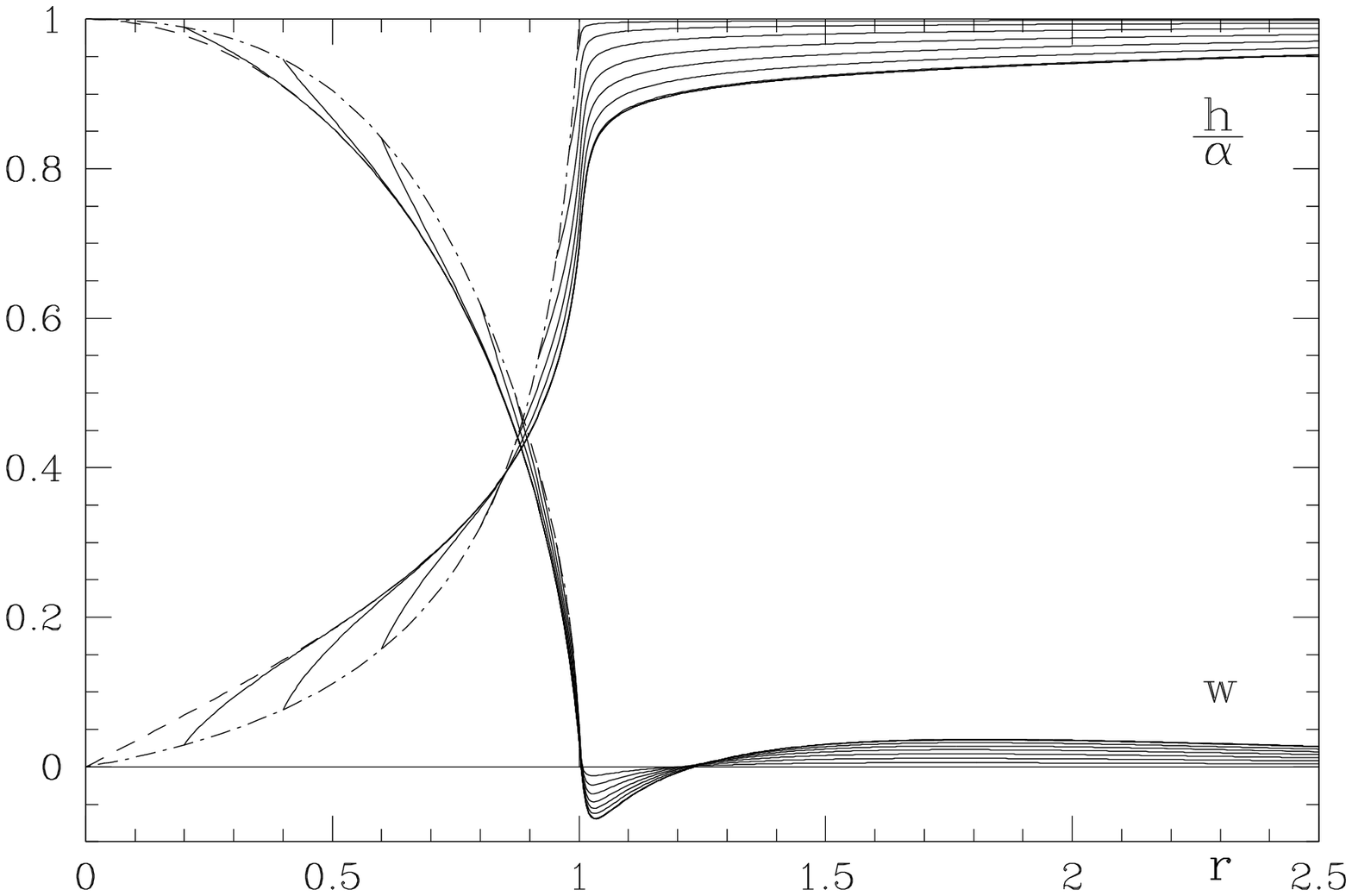,width=0.45\linewidth}\hss
  \epsfig{bbllx=41bp,bblly=178bp,bburx=576bp,bbury=517bp,%
  file=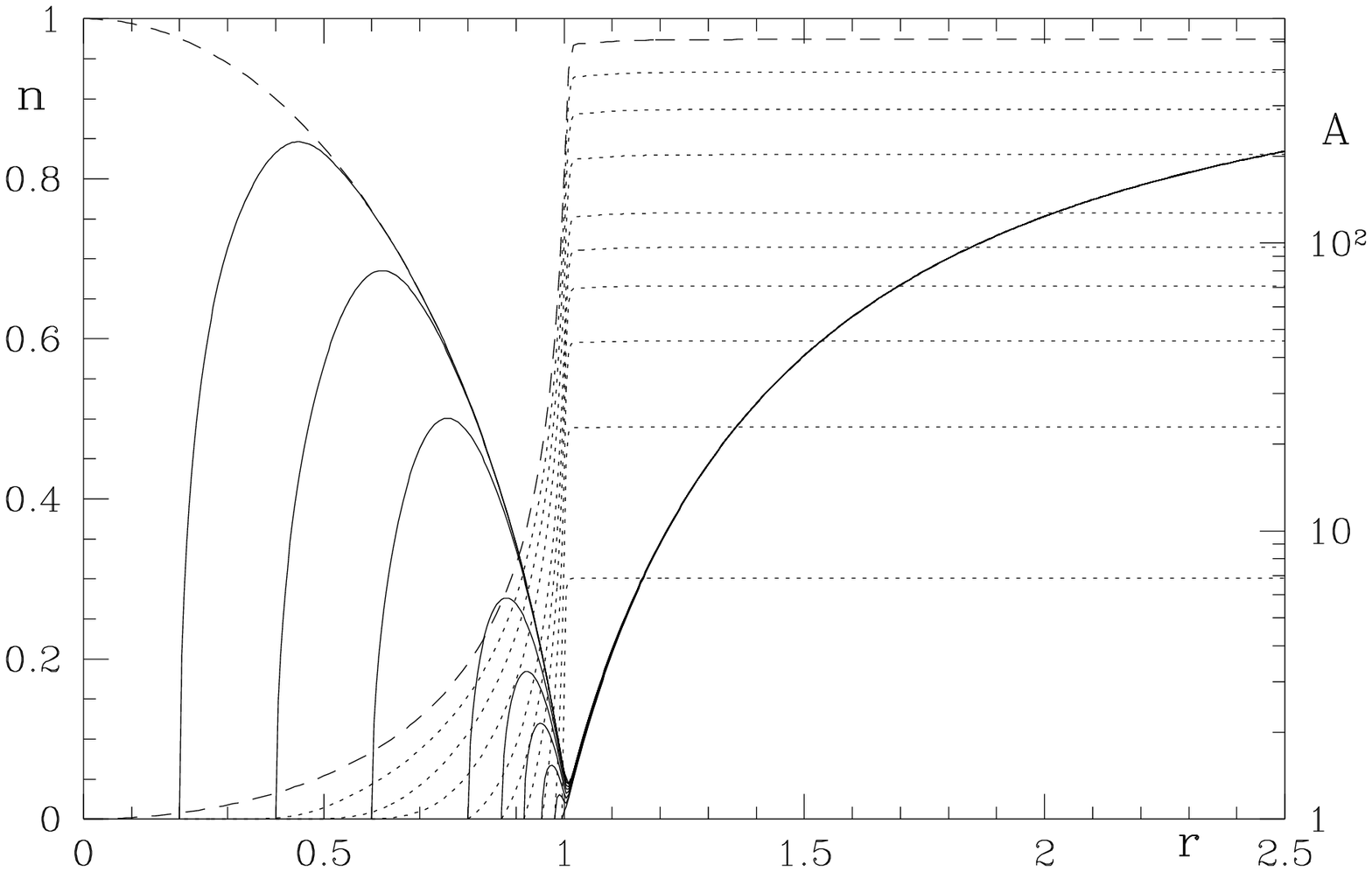,width=0.45\linewidth}\hss
}\vspace{-0.3cm}
\caption[figh2a040p4d16]{\label{figh2a040p4d16}Excited, $N=2$ black monopole
solutions for $p=4$ and $\alpha=0.4$, with $r_h=0.2$, 0.4, 0.6, 0.8, and
$w_h=0.5$, 0.4, 0.3, 0.2, and 0.1. The dashed lines are for the
corresponding regular monopole. The dashed-dotted lines show the
$r_h$-dependence of $h_h$ and $w_h$.}
\end{figure}
The initial data are quite similar for different values of $p$ and are shown
in Fig.~\ref{figphwh} for $p=3$ with $r_h=0.5$ and for $p=4$ with $r_h=1.5$.
For $h_h\ll1$, and thus $\alpha\ll1$ they start with $w_h\approx1$ for the
fundamental, $N=0$ solution (tiny Schwarzschild black hole inside a flat
space BPS monopole), or with $w_h$ as for the corresponding EYM black hole,
$N=1$, 2,~$\ldots$, or $N\to\infty$ if $r_h<1$ (compare Fig.~3
of~\cite{Breitenlohner:1993es}).

For $r_h<1$ the families end with critical solutions as for the regular
monopoles. The critical fundamental solutions occur at a value $\alpha_{\rm
cr}(r_h)$, starting for $r_h=0$ from $\alpha_{\rm cr}$ as given in
Tab.~\ref{tabMalpha} for the regular monopoles and reaching the value
$\sqrt{0.75}$ for some $r_h$ close to but less than~1. For larger $r_h$ the
fundamental solutions converge to the limiting one as $\alpha^2\to0.75$ and
cease to exist beyond that value, as do the excited solutions for all
$r_h<1$.

The situation is quite different for $r_h>1$. All solutions end with $w_h=0$
where they bifurcate with a regular, {\it i.e.}, non-extremal RN black hole. Black
monopole solutions with various values of $r_h$ are shown for $p=4$, $N=2$,
$\alpha=0.4$ in Fig.~\ref{figh2a040p4d16}.

The relatively simple domain of existence of the families of black monopoles
described above gets rather more complicated by the existence of
maxima and minima (of $\alpha$ for fixed $r_h$ or vice versa), giving rise
to various branches.
\begin{figure}[ht]
\hbox to\linewidth{\hss
  \epsfig{bbllx=40bp,bblly=183bp,bburx=561bp,bbury=517bp,%
  file=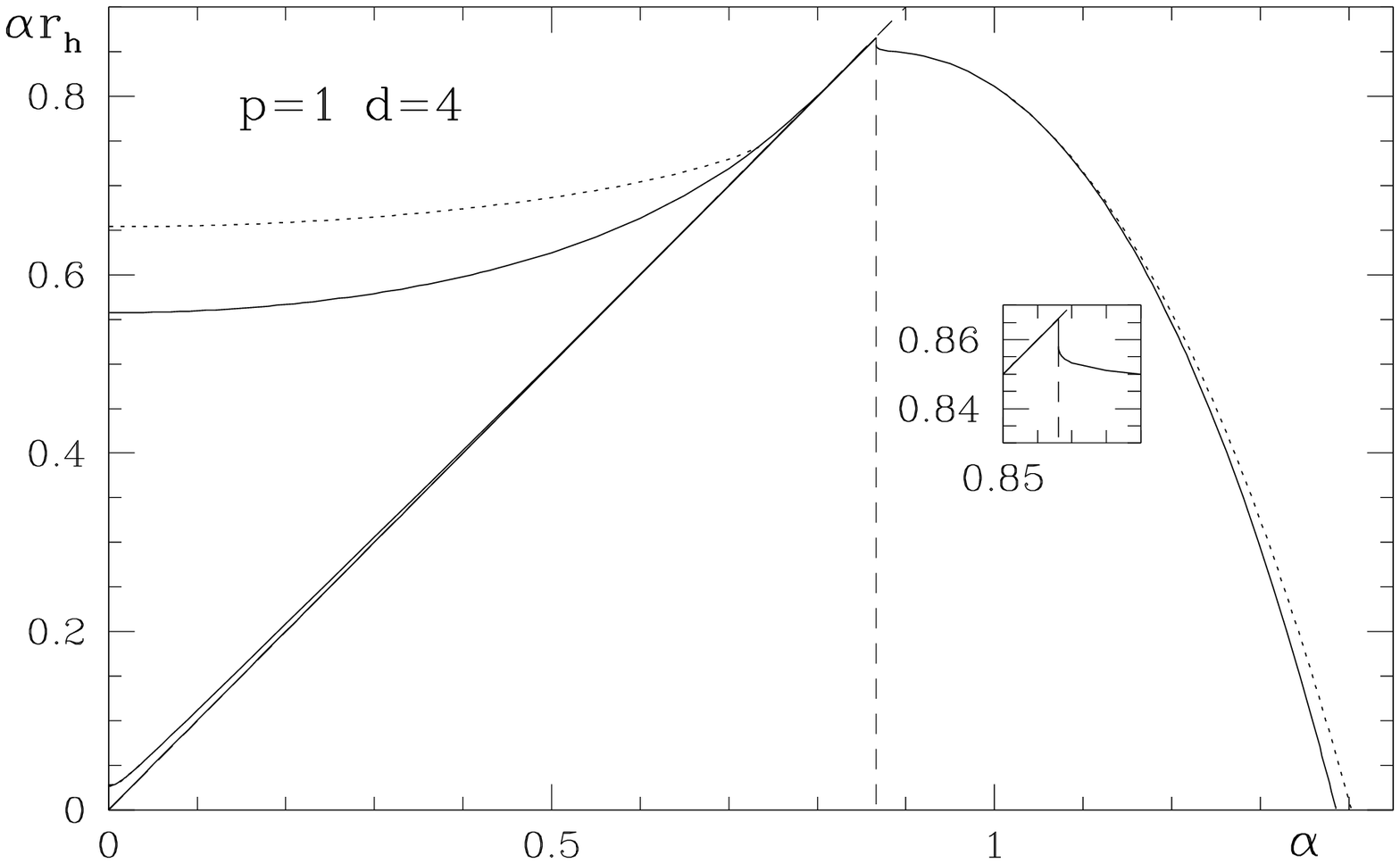,width=0.45\linewidth}\hss
  \epsfig{bbllx=40bp,bblly=183bp,bburx=561bp,bbury=517bp,%
  file=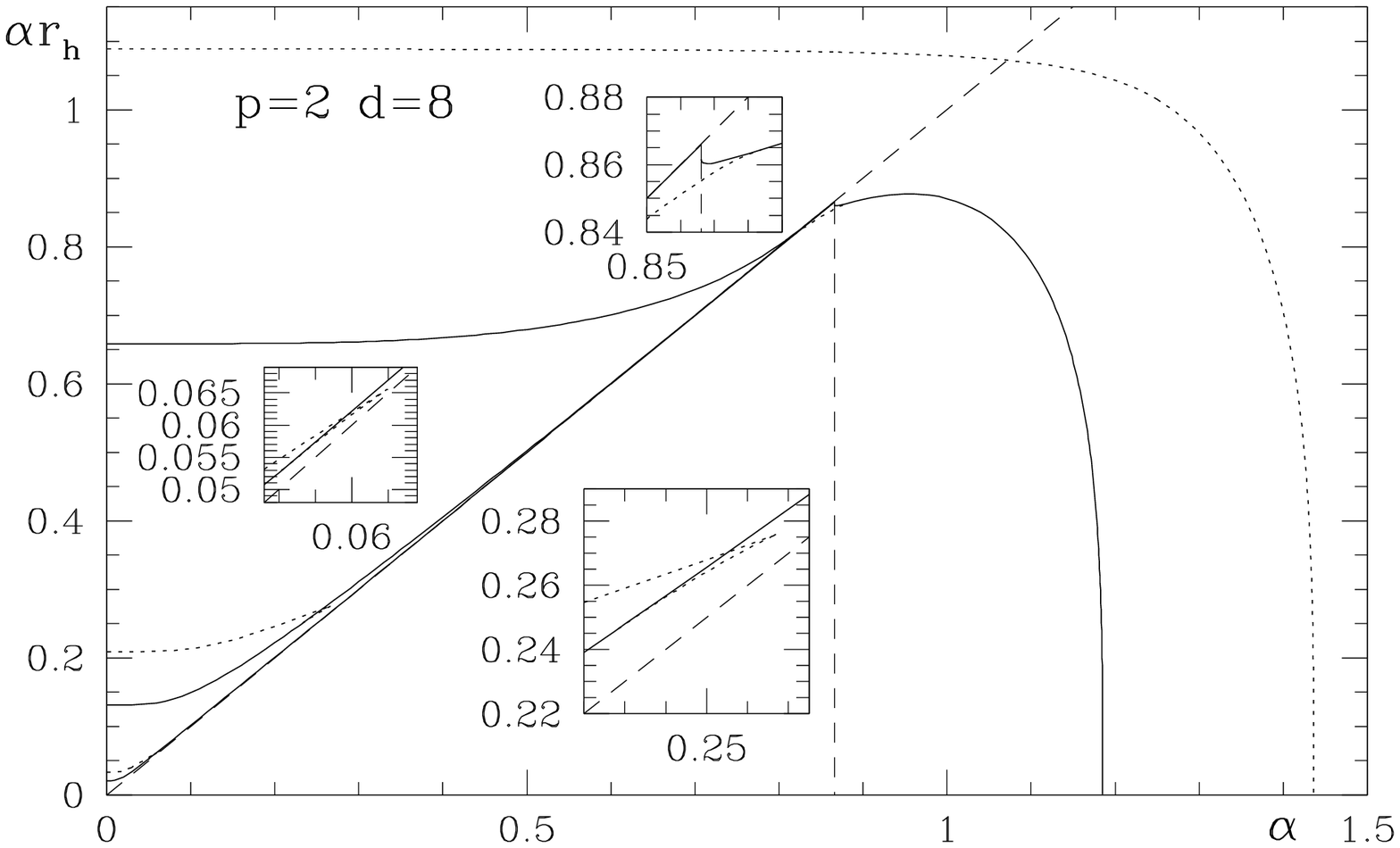,width=0.45\linewidth}\hss
}\vspace{0.2cm}
\hbox to\linewidth{\hss
  \epsfig{bbllx=40bp,bblly=183bp,bburx=561bp,bbury=517bp,%
  file=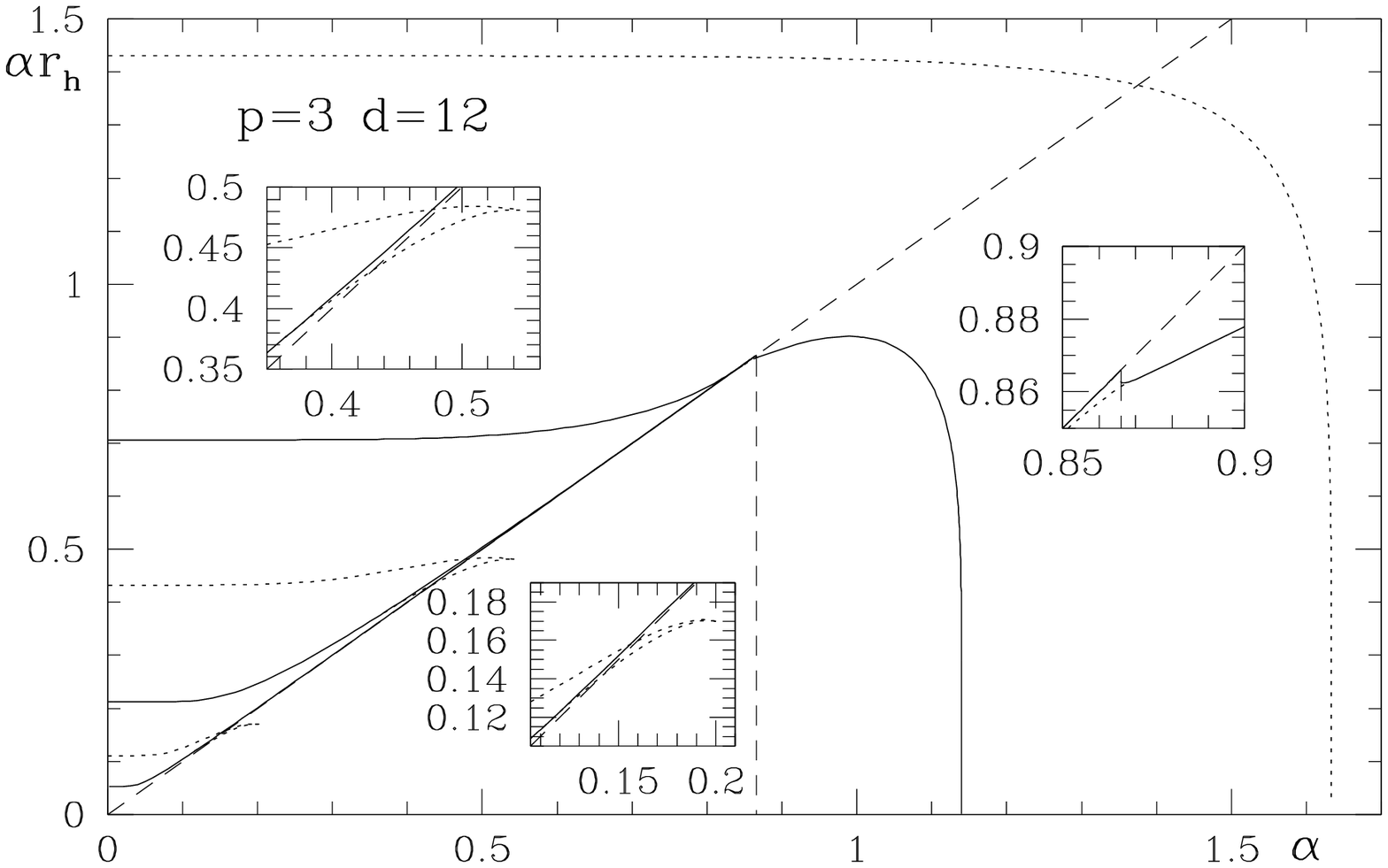,width=0.45\linewidth}\hss
  \epsfig{bbllx=40bp,bblly=183bp,bburx=561bp,bbury=517bp,%
  file=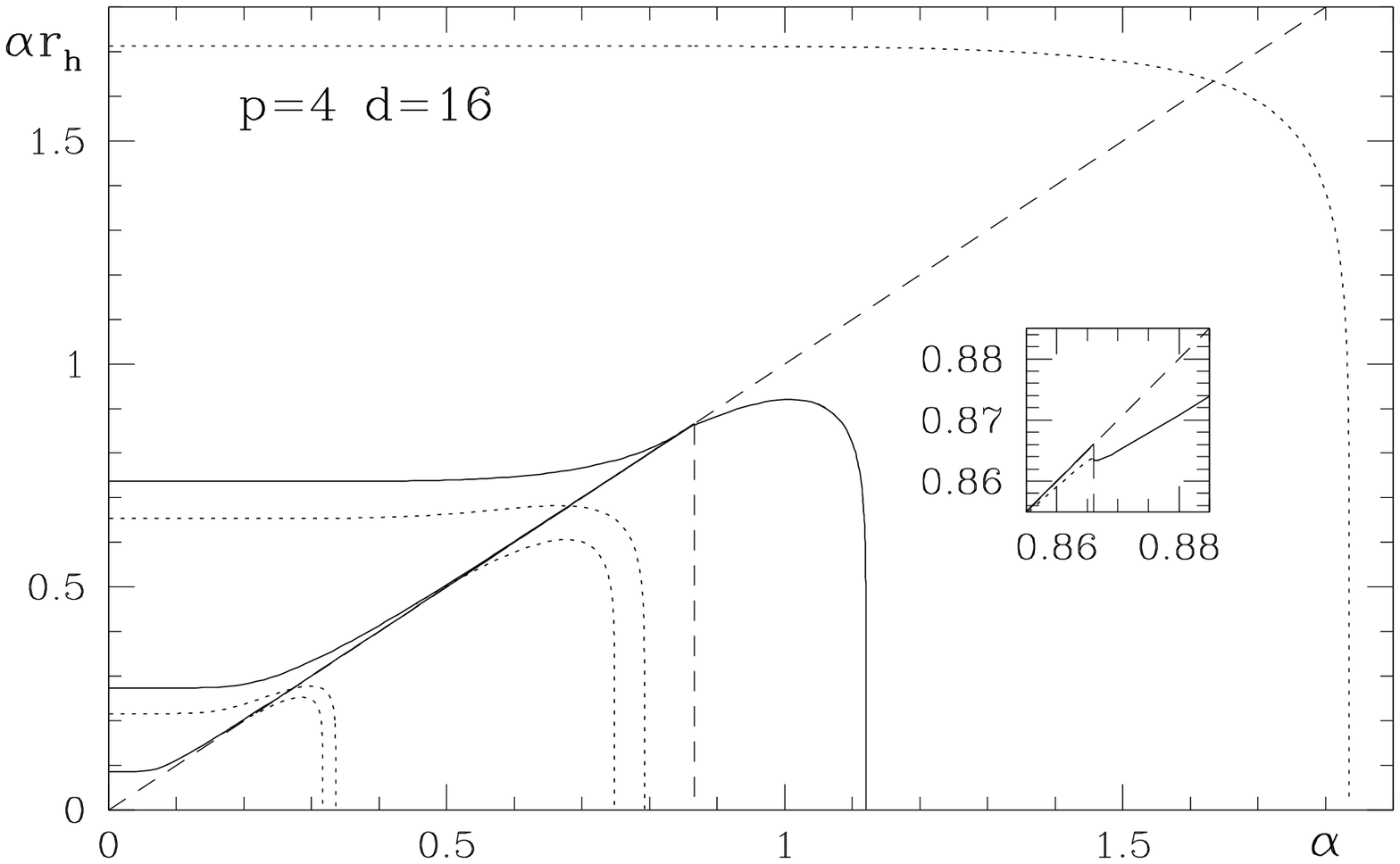,width=0.45\linewidth}\hss
}\vspace{-0.3cm}
\caption[figdomain]{\label{figdomain}Domains of existence for (fundamental
and excited) black monopole solutions with $p=1$, 2, 3, and~4.}
\end{figure}
Fig.~\ref{figdomain} shows these domains as $\alpha r_h$ {\it vs.}\ $\alpha$ for
$p=1$, 2, 3, and~4. Regular RN black holes only exist for $r_h>1$ (above the
dashed diagonal). Fundamental and excited black monopoles always exist for
$r_h<1$ (below the diagonal) and $\alpha^2<0.75$ (left of the dashed
vertical line), but extend beyond that region. Exhibiting maxima and minima
(dotted lines) they either end as critical solutions (solid lines for
$\alpha^2\ge0.75$, including a short vertical line near $r_h=1$) or
bifurcate with a regular RN black hole (solid lines for $\alpha^2<0.75$).

The fundamental solutions with $r_h\ll1$ always have two branches, a `lower'
one from $\alpha=0$ to some $\alpha_{\rm max}(r_h)$ and an `upper' one from
$\alpha_{\rm max}(r_h)$ to $\alpha_{\rm cr}(r_h)$. Likewise, the fundamental
and excited solutions with $\alpha\ll1$ always have two branches, a lower
one from $r_h=0$ to some $r_{\rm max}(\alpha)$ and an upper one from from
$r_{\rm max}(\alpha)$ to $r_{\rm bif}(\alpha)$.

For $p=1$ these upper branches of the fundamental solution cease to
exist near $r_h=1$. For $p>1$ they exist for all $r_h$ and are connected.
In these cases continuity requires the existence of yet another branch near
$r_h$ and $\alpha=\sqrt{0.75}$, shown as detail in Fig.~\ref{figdomain}.

\begin{figure}[ht]
\hbox to\linewidth{\hss
  \epsfig{bbllx=40bp,bblly=183bp,bburx=561bp,bbury=517bp,%
  file=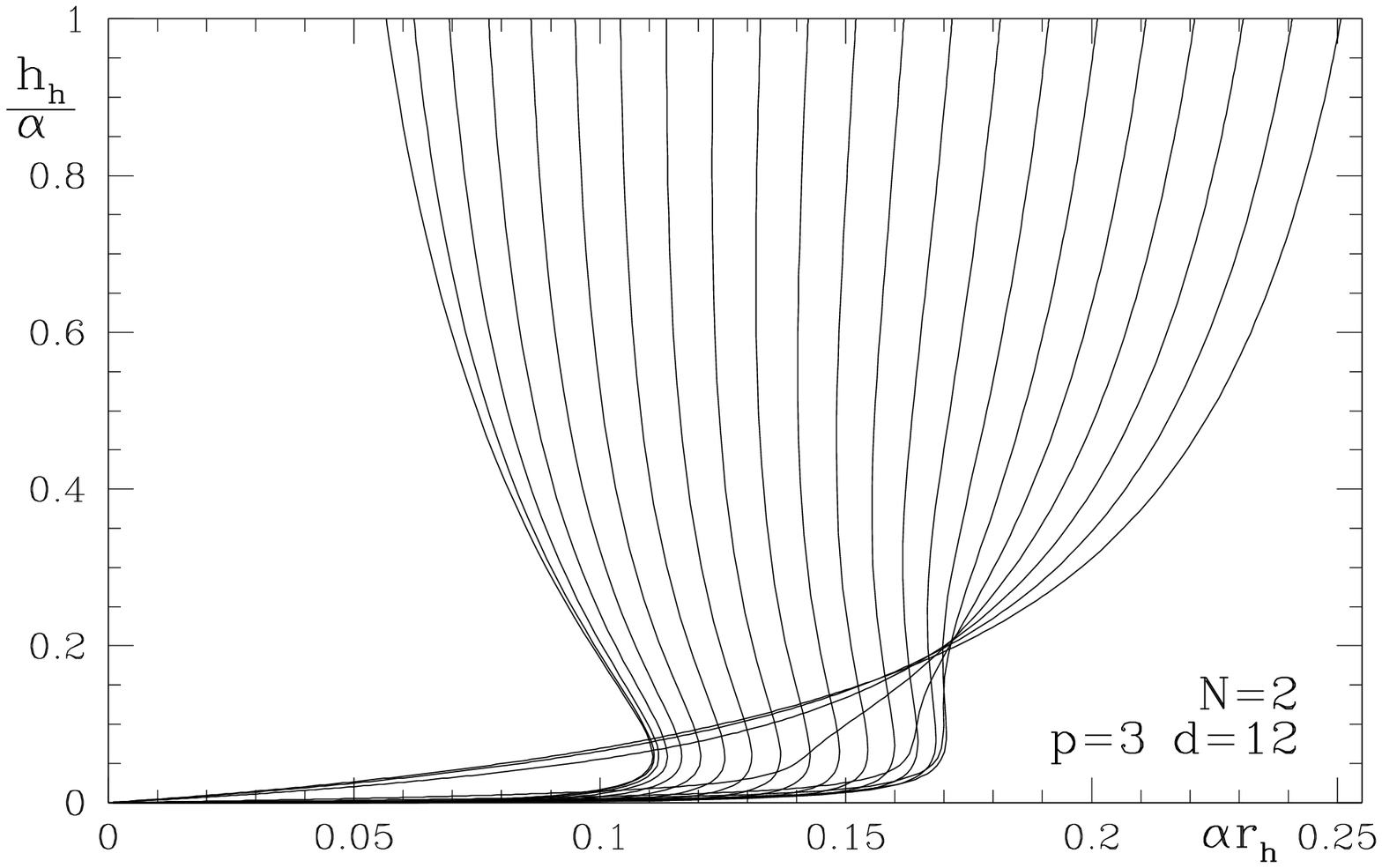,width=0.45\linewidth}\hss
  \epsfig{bbllx=40bp,bblly=183bp,bburx=561bp,bbury=517bp,%
  file=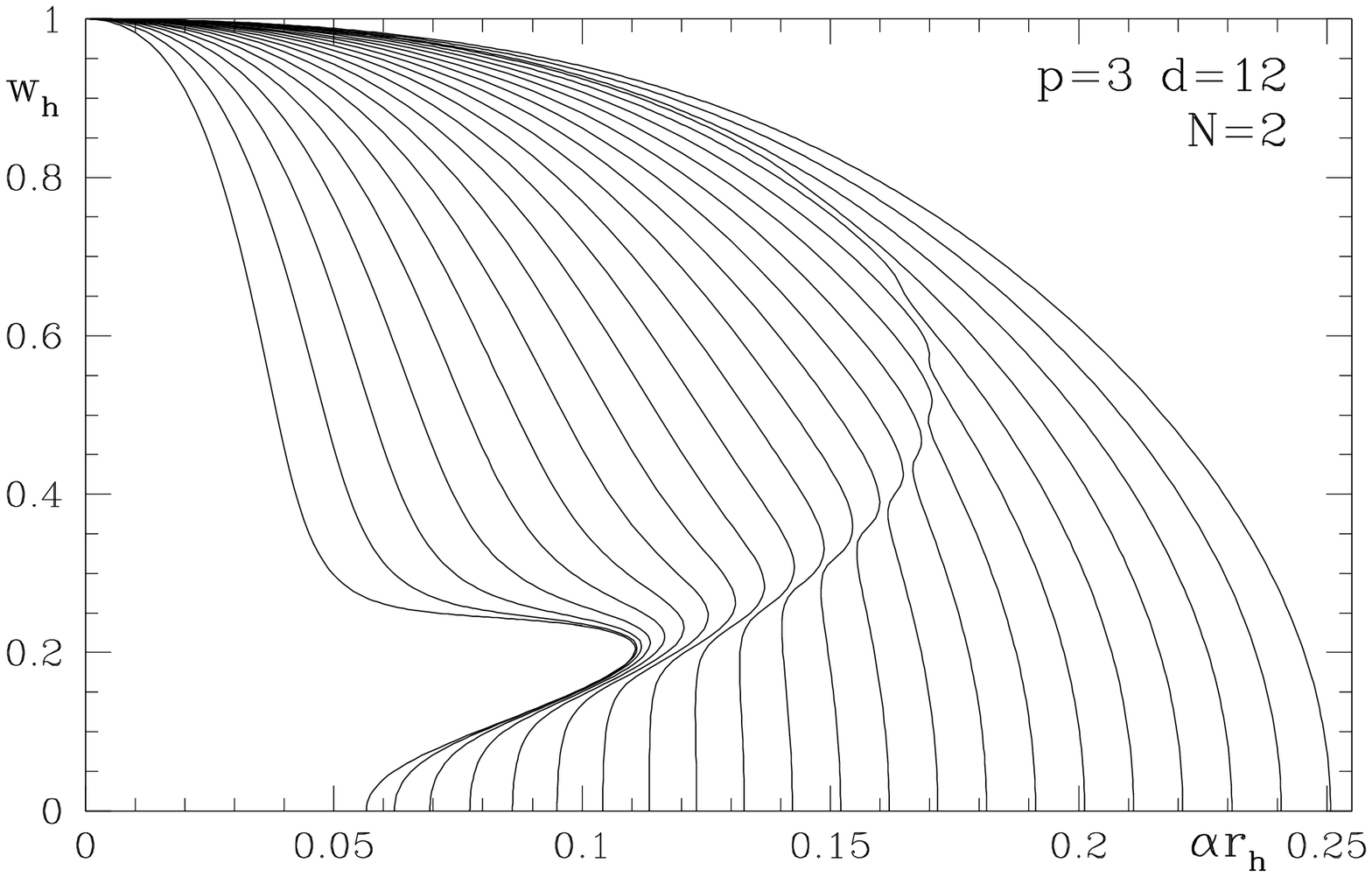,width=0.45\linewidth}\hss
}\vspace{-0.3cm}
\caption[figrhphwh2p3d12]{\label{figrhphwh2p3d12}Initial data for the $N=2$
excited black monopole solutions for $p=3$ with $\alpha=0.04$, 0.05,
$\ldots$,~0.25.}
\end{figure}
For $p=1$ the upper branch of the excited solutions ceases to exist for
larger values of $\alpha$. For $p>1$ the situation is somewhat more
complicated. Increasing $\alpha$, there first appears a minimum of $r_h$
near the bifurcation, giving rise to a third branch. For $p=2$ and~3 the
maximum and minimum eventually meet, {\it i.e.}, the upper branch disappears and
the lower and third branch are joined (compare Fig.~\ref{figrhphwh2p3d12}).
For $p=4$, however, all three branches continue as $r_h\to0$, with the
`upper' branch of the black monopoles corresponding to the `middle' one of
the regular monopoles. This correspondence suggests that the third branch
observed for $p=2$ and~3 is related to the steep (almost vertical) increase
of $b$ {\it vs.}\ $\alpha$ shown in Fig.~\ref{figbalpha}.

\section{Summary and Discussion}
When it comes to choosing a model for a gravitating monopole in higher
dimensions, the number of options proliferate.  Our choices are guided by the twin criteria of
having a Bogomol'nyi lower bound that can be saturated in the flat space
limit of the Yang-Mills--Higgs (YMH) subsystem, and that aside from
the gravitational coupling and the Higgs VEV there feature no other
dimensionful constants in the model.  This restriction is made to simplify
both the analytic and the numerical analyses.  The YMH systems employed, in
$d=4p$ spacetime, are those resulting from dimensional descent from
$R^{4p-1}\times S^1$ which in flat space support selfdual
solutions~\cite{Radu:2005rf}.  The gravitational systems are then chosen to
be those members of the gravitational hierarchy (\ref{gp}) with $p=q$, such
that the relation between the dimensions of the gravitational term in the
Lagrangian (\ref{lag}) has exactly the same relation to the dimensions of
the YMH term, for all $p$.

We have carried out a precise quantitative analysis of gravitating static
monopoles, both regular and black, in $d=4p$ spacetime dimensions.  The
models we have employed are in a sense very special, but the properties of
their solutions are generic.  Our template has been the gravitating
Georgi-Glashow model in $d=4$, in the BPS limit, {\it i.e.}, in the absence
of the Higgs self interaction potential.  This was the part of the subject
of study in \cite{Breitenlohner:1991aa, Breitenlohner:1994di} in which the
Higgs potential was absent.  We have emulated the results of
\cite{Breitenlohner:1991aa, Breitenlohner:1994di} exactly here, and have
thus achieved our aim of showing that the generic properties of gravitating
monopoles are fully understood at least within the restricted choice of
models that we have exercised.  As such, the present study is a general
preliminary investigation into the nature of gravitating monopoles in higher
dimensions, both regular and black.

In Section~2, we have presented an exhausive analytic analysis of the
residual one dimensional action of the static system subject to spherical
symmetry in the spacelike dimensions.  Especially prominent is the analysis
of the Reissner--Nordstr\"om fixed point, and due to our restricted choice
of systems, no {\em conical fixed points}~\cite{Breitenlohner:2005hx}
feature here.  (Some numerical analysis of less restricted models was
carried out which indicated the existence of conical fixed points, but we do
not report on these here.) A preliminary step in the analysis
for the gravitating Yang--Mills $p-$hierarchy was carried out before
proceeding to the case of YMH in $d=4p$.  This underpins the numerical
results of \cite{Radu:2006mb}.  That was followed in Section~3 by the
numerical analysis of the $4p-$dimensional gravitating monopoles, where the
qualitative features discovered in \cite{Breitenlohner:1991aa,
Breitenlohner:1994di} for the $p=1$ case were reproduced quantitatively for
$p=1,2,3$ with high accuracy.  It is clear that the regular pattern observed
{\it modulo} $4p$ up to $p=3$ will repeat for all $p$.

Having achieved our aim of exhibiting the regularity of features {\it
modulo} $4p$ of the simplest class of gravitating monopoles, it is perhaps
in order to point to natural succeeding investigations.  In this context we
would exclude the otherwise obvious choice of employinging the usual ($p=1$)
Einstein--Hilbert system in models in $d\ge 5$, since this would result in
the additional complication of encountering conical fixed points.  As for
introducing the usual Yang--Mills system in $d\ge 5$, this is excluded on
the grounds that already in the flat space limit it would cause the
mass/energy to diverge because of the {\it half--pure-gauge} asymptotic
decay of the YM connection in all dimensions.  The following options are
open.

{\bf a)} Extend the static spherically symmetric Ansatz (\ref{connection}) to
allow for the electric YM connection
$A_0=u(r)\,\frac{x_j}{r}\,\Sigma_{j\,D-1}$.  These would yield the
gravitating versions of the dyons constructed in \cite{Radu:2005rf}.

{\bf b)} Construct the gravitating monopoles in {\em odd} spacetime dimensions. 
The main difference of the YMH models to be employed in this task, and the
ones used in the present work, is that unlike the latter the former arise
from the dimensional descent from $R^D \times K^N$ where $N$ is now even and
hence $N\ge 2$.  As such, the gravity decoupling limits of these solutions
do not saturate the Bogomol'nyi lower bound~\cite{Tchrakian:1990gc}, but
this is not important since in any case the gravitating monopoles are not
selfdual.  In this respect gravitating monopoles in {\em odd} spacetimes are
similar to those in {\em even} spacetimes for $N\ge 3$, with odd $N$, so
that the two tasks should be performed in parallel.  Another difference
between $N=1$ and $N>1$ models is that in the $N>1$ models expressions of
the Lagrangians for increasing $p$ become progressively more cumbersome, in
contrast with the $N=1$ models for which the $p-$hierarchy is quite uniform. 
For this reason in the $N>1$ case it is reasonable to restrict to $p=2$ to
get a glimpse of the qualitative features.  Were one to restrict attention
to the $p=2$ YMH systems, then presumably the most aesthetic (if not
necessary) choice for the gravitational system would be the $p=2$ member of
the gravitational hierarchy.  Restricting to $p=2$ YMH systems, the
pertinent examples are $d=6+1$, $d=5+1$ and $d=4+1$ spacetimes.  The $p=2$,
$d=3+1$ case is also of some interest, as it presents very different
properties from the usual $p=1$, $d=3+1$ case, notably supporting mutually
attracting like-charged monopoles~\cite{Kleihaus:1998gy}.

\subsection*{Acknowledgements}
It is a pleasure to thank Dieter Maison and Eugen Radu for their
participation at the early stages.  This work is supported in part by
Science Foundation Ireland (SFI) in the framework of project RFP07-330PHY.

\appendix
\section{The New Variables}
With the substitutions Eqs.~(\ref{tausubst}) the Action~(\ref{reducedactionGM})
takes the form
\bsea\label{Stau}
S_G&=&-\frac{1}{2p}\int d\tau\,e^{\nu+\lambda}\frac{1}{\dot r}\frac{d}{d\tau}
  \Bigl(r^{d-2p-1}(1-e^{-2\lambda}\dot r^2)^p\Bigr)\;,
\\
S_M&=&\int d\tau\,e^{\nu+\lambda}\,r^{d-4p}\Biggl[W^{p-1}
  \Bigl(e^{-\lambda}\frac{dw}{d\tau}\Bigr)^2
  +\frac{d-2p-1}{2p}\frac{W^p}{r^2}
\nonumber\\&&\qquad\qquad\qquad\quad
  +\frac{1}{2p}\Bigl(re^{-\lambda}\frac{dH_p}{d\tau}\Bigr)^2
  +(d-2p-1)w^2H_p^2\Biggr]\;.
\esea
As a first step we want to absorb all $\tau$-derivatives of $\lambda$ in
Eq.~(\ref{Stau}a) into a surface term.  To write this in compact form
we introduce a polynomial $F_p$ uniquely determined by the property
$F_p(z)+2zF_p'(z)=(1-z)^p$. We thus obtain
\bea\label{SGtau1}
S_G&=&\int d\Phi(e^{-\lambda}\dot r)
  -\int d\tau\,e^{\nu+\lambda}\,r^{d-2p-2}\Biggl[
  (re^{-\lambda}\dot\nu)(e^{-\lambda}\dot r)
    F_p(e^{-2\lambda}\dot r^2)
\nonumber\\&&\qquad
  +(d-2p-1)\Bigl[\frac{1}{2p}(1-e^{-2\lambda}\dot r^2)^p
    +(e^{-\lambda}\dot r)^2
      F_p(e^{-2\lambda}\dot r^2)\Bigr]\Biggr]\;,\qquad
\eea
with
\be
\Phi(x)=e^\nu r^{d-2p-1}x\,F_p(x^2)\;.
\ee

Next, we consider an action of the form
\be\label{action}
S=\int d\tau\,e^\lambda\,L(\varphi,e^{-\lambda}\dot\varphi)\;,
\ee
and introduce $\psi_i$ as {\em abbreviation} for
$e^{-\lambda}\dot\varphi_i$.  Varying this action w.r.t.\ $\varphi_i$ yields
the second order equations
\be\label{actionsecond}
\sum_j M_{ij}\,e^{-\lambda}\dot\psi_j=
  \frac{\partial L}{\partial\varphi_i}
  -\sum_j\frac{\partial^2L}{\partial\psi_i\partial\varphi_j}
   \psi_j\;,
\qquad{\rm where}\quad
  M_{ij}=\frac{\partial^2L}{\partial\psi_i\partial\psi_j}\;,
\ee
and we assume that the determinant $|M_{ij}|$ does not vanish identically.
Varying the action w.r.t.\ $\lambda$ yields the first order equation
\be\label{actionconstr}
L=\sum_i\psi_i\frac{\partial L}{\partial\psi_i}\psi_i\;,
\ee
{\it i.e.}, the reparametisation constraint.

In order to obtain a system of first order differential equations we now
want to introduce {\em new variables} $\psi_i$ as Lagrange multipliers.  For
a Lagrangian of the form
\be
L(\varphi,\psi)=P(\varphi)+\sum_{ij}\psi_i\psi_j\,K_{ij}(\varphi)\;,
\ee
{\it i.e.}, qudratic in the derivatives, this is achieved by the replacement
\bea
L(\varphi,e^{-\lambda}\dot\varphi)\Rightarrow
  \tilde L&=&L-\sum_{ij}(e^{-\lambda}\dot\varphi_i-\psi_i)
  (e^{-\lambda}\dot\varphi_j-\psi_j)\,K_{ij}
\nonumber\\
&=&P+\sum_{ij}(e^{-\lambda}\dot\varphi_i\psi_j
 +\psi_ie^{-\lambda}\dot\varphi_j-\psi_i\psi_j)K_{ij}\;.
\eea
We can generalise this procedure to the case with higher powers of
derivatives by the replacement
\be
L(\varphi,e^{-\lambda}\dot\varphi)\Rightarrow\tilde L=
  L(\varphi,\psi)+\sum_i(e^{-\lambda}\dot\varphi_i-\psi_i)
  \frac{\partial L(\varphi,\psi)}{\partial\psi_i}\;.
\ee
Varying the new action $\tilde S=\int d\tau\,e^\lambda\,\tilde L$ w.r.t.\
the new variables $\psi_i$ yields
\be
\sum_j M_{ij}(e^{-\lambda}\dot\varphi_j-\psi_j)=0\;,
\ee
{\it i.e.}, what used to be abbreviations are now field equations.
Varying w.r.t.\ $\varphi_i$ or $\lambda$ and substituting these new
field equations reproduces the previous results Eqs.~(\ref{actionsecond})
and ~(\ref{actionconstr}).

Applying this procedure to $\left(S_G-\int d\Phi\right)$ in Eq.~(\ref{SGtau1}) and
replacing $\Phi(e^{-\lambda}\dot r)$ by $\Phi(n)$ yields Eq.~(\ref{Stau3}a).
Applying the procedure to Eq.~(\ref{Stau}b) yields Eq.~(\ref{Stau3}b).



\end{document}